\documentclass[%
aps,authoryear
]{aastex62}
\usepackage{graphicx}
\usepackage{ulem}
\usepackage{bm}
\usepackage{color}
\usepackage{textcomp}
\usepackage{gensymb}
\usepackage{amssymb,amsmath}
\shorttitle{PSP context: critical surfaces}
\shortauthors{Chhiber et al.}

\newcommand{\rs}{R_\odot}
\newcommand{\dash}{\text{ -- }}
\newcommand{\bb}{\mathbf{B}}
\newcommand{\vb}{\mathbf{v}}
\newcommand{\psp}{\textit{PSP} }

\newcommand{\varepsilonb}{\mbox{\boldmath$\varepsilon$}}

\newcommand{\Rb}{\mbox{\boldmath$\cal R$}}

\begin{document}
\title{Contextual Predictions for Parker Solar Probe I: Critical Surfaces and Regions}
\author{Rohit Chhiber}
\email{rohitc@udel.edu}
\affiliation{Department of Physics and Astronomy, University of Delaware, Newark, DE 19716, USA}
\author{Arcadi V.~Usmanov}
\affiliation{Department of Physics and Astronomy, University of Delaware, Newark, DE 19716, USA}
\affiliation{NASA Goddard Space Flight Center, Greenbelt, MD 20771, USA} 
\author{William H.~Matthaeus}
\affiliation{Department of Physics and Astronomy, University of Delaware, Newark, DE 19716, USA}
\author{Melvyn L.~Goldstein}
\affiliation{NASA Goddard Space Flight Center, Greenbelt, MD 20771, USA} 
\affiliation{University of Maryland Baltimore County, Baltimore, MD 21250, USA}
\begin{abstract}
The solar corona and young solar wind may be characterized by \textit{critical surfaces} -- the sonic, Alfv\'en, and first plasma-$\beta$ unity surfaces -- that demarcate regions where the solar wind flow undergoes certain crucial transformations. Global numerical simulations and remote sensing observations offer a natural mode for the study of these surfaces at large scales, thus providing valuable context for the high-resolution in-situ measurements expected from the recently launched \textit{Parker Solar Probe} (\textit{PSP}). The present study utilizes global three-dimensional magnetohydrodynamic (MHD) simulations of the solar wind to characterize the critical surfaces and investigate the flow in propinquitous regions. 
Effects of solar activity are incorporated by varying source magnetic dipole tilts and employing magnetogram-based boundary conditions. An MHD turbulence model is self-consistently coupled to the bulk-flow equations, enabling investigation of turbulence properties of the flow in the vicinity of critical regions. The simulation results are compared with a variety of remote sensing observations. A simulated \psp trajectory is used to provide contextual predictions for the spacecraft in terms of the computed critical surfaces.
Broad agreement is seen in the interpretation of the present results 
in comparison with existing remote sensing results, both from heliospheric 
imaging and from radio scintillation studies. 
The trajectory analyses show that the period of time that \psp is 
likely to spend inside the $\beta=1$, sonic, and Alfv\'en surfaces 
depends sensitively on the degree of solar activity and the tilt of 
the solar dipole and location of the heliospheric current sheet. 
\end{abstract}
\keywords{solar wind --- Sun: corona --- magnetohydrodynamics (MHD) --- turbulence}
\section{Introduction}\label{sec:int}
The expansion of the solar corona into interplanetary space was predicted in 1958 by Parker's classic model \citep{parker1958apj}. Soon after, in-situ spacecraft measurements \citep{neugebauer1966jgr} confirmed that the interplanetary region is pervaded by solar plasma flowing at supersonic speed.\footnote{For a recent historical review of the discovery of the solar wind, see \cite{obridko2017SoSyR}.} Research efforts in the following decades have established that the solar wind is a complex and dynamic system  
that enters centrally into much of space research 
and is of relevance to studies of solar, geophysical, and astronomical phenomena. 
The \textit{Parker Solar Probe} (\textit{PSP}) mission \citep{fox2016SSR} \replaced{is scheduled for a Summer 2018 launch}{was launched on August 12, 2018}, 
with the goal of exploring, for the first time,
regions of solar wind that are of crucial importance 
in establishing the heliosphere. 
While approaching the Sun closer than any prior spacecraft,
\psp will provide 
unprecedented high-resolution measurements of the 
solar corona and the young solar wind, with \replaced{its}{the} main objectives
being discovery of 
the structure and dynamics of the coronal magnetic field,  
and the processes that 
heat and accelerate the wind  
and accelerate and transport energetic particles. 
As the \psp makes its high-resolution in-situ measurements, 
a knowledge of the large-scale environment within which 
these observations exist is of vital importance.
This global context may be provided by remote sensing \citep{bird1990POIH,vourlidas2016ssr} 
and global simulation. The present work is the first of a 
series of papers focused on contextual predictions 
for \psp using global simulations of the solar wind.

\replaced{The transition of the solar corona into the solar wind 
is accomplished by several dynamical changes in the 
nature of the flow. The inner corona is magnetically structured, subsonic, and sub-Alfv\'enic, but as the solar plasma flows out from the corona into the young solar wind, it transforms into a supersonic and super-Alfv\'enic flow that is dominated by hydrodynamics.}{The transition of the solar corona into the solar wind is accomplished by several dynamical changes in the nature of the flow, regionally organized by magnetic topology and associated factors such as open vs. closed connectivity and composition. Regions of fast wind, slow wind, and mixed wind apparently trace to different magnetic connectivities and different altitudes \citep[e.g.,][]{mccomas2003grl,cranmer2007ApJS}. In the simplest picture the inner-coronal plasma is magnetically structured, subsonic, and sub-Alfvenic, but as it flows out from the corona into the young solar wind it evolves into a supersonic and super-Alfvenic flow that is dominated by hydrodynamics.} Recent work indicates that this transition 
may coincide with the onset of large-scale turbulence 
\citep{deforest2016ApJ828,chhiber2018apjl} 
and mark the outer boundary of a zone of 
preferential ion heating \citep{kasper2017apj}.

Useful markers that characterize this transition 
are the sonic critical surface, 
the Alfv\'en critical surface, 
and the first $\beta = 1$ surface (the plasma-$\beta$ is 
the ratio of gas to magnetic pressure). 
In particular, when the flow speed $u$ 
exceeds the Alfv\'en speed $V_A$, the magnetic field rigidity can 
no longer enforce plasma co-rotation \citep{weber1967ApJ148}, or overcome the differential ram-pressure due to shearing interactions between neighbouring wind streams. 
And when the plasma-$\beta$ increases above unity, gradients in the plasma (thermal)
pressure may displace the magnetic field and more isotropic 
motions are possible \citep{chhiber2018apjl}. \replaced{The region in which 
these two crucial conditions ($u > V_A$ and $\beta \sim 1$) are 
attained becomes, in effect, the region where the 
corona gives up control of the solar plasma, 
and the solar wind as an independent entitity is born.}{The broad region in which these two crucial conditions \((u > V_A\) and \(\beta\sim 1)\) are attained becomes, in effect, the region where the corona gradually gives up control of the solar plasma, and the kinetic-energy dominated solar wind emerges as an independent entity. Beyond these regions the solar wind no longer communicates through magnetohydrodynamic (MHD) interactions with the magnetically dominated regions of its origin.} 

In this work we employ well-tested global MHD simulations of the solar wind \citep{usmanov2014three,usmanov2018}, that are self-consistently 
coupled with a turbulence transport model, to study and characterize this 
region of transitions and to make contextual predictions for the \psp mission.\footnote{Our use of ``transition'' here 
should not be confused with the well-known transition region that lies 
just above the chromosphere \citep[e.g.,][]{cranmer2007ApJS}.}  We incorporate the effects of long-term solar variability 
\citep[e.g.,][]{owens2013lrsp} by varying magnetic source 
dipole tilts and employing magnetogram-based boundary conditions. 
The simulation results are compared with a variety of remote sensing observations, 
demonstrating how the two approaches may be combined to gain 
insights regarding large scale heliospheric conditions in this region. 
Global simulation and remote sensing thus generate mutual support, 
and in turn, provide valuable context for the finer details that 
emerge from in-situ measurements. Subsequent papers in this series on 
contextual predictions for \psp will focus on turbulence properties 
along the spacecraft's trajectory, on modifications of Taylor's hypothesis 
for \psp \citep{matthaeus1997AIPCtrajectory,klein2015ApJtaylor}, and on  
solar wind azimuthal flow.

The paper is organized as follows -- in Section \ref{sec:back} we 
provide background on critical surfaces 
and physically distinct regions of the inner wind, 
discussing recent work that motivates the present study. 
An overview of the \psp trajectory is provided in Section \ref{sec:sampl}, and our solar wind model is briefly described in Section \ref{sec:model}. 
Results are presented in Section \ref{sec:results}, including 
comparisons of model output with remote sensing observations 
and contextual predictions along the \psp trajectory.
We conclude with discussion in Section \ref{sec:disc}.
\section{Theoretical and Observational Background}\label{sec:back}
%

Two critical points\footnote{A mathematical discussion of a critical (or equilibrium) point of a system of ordinary differential equations may be found in standard texts \citep[e.g.,][]{boyce1969elementary}.} are frequently discussed within the context of the solar wind -- the sonic and the Alfv\'enic critical points, where the flow speed equals the sound speed and the Alfv\'en speed, respectively. One encounters the notion in even the simplest, spherically symmetric, stationary and isothermal model of the solar wind \citep[e.g.,][]{hundhausen1972coronal}. We briefly review the standard presentation below. 

The relevant equations may be derived by assuming an equal number density $n$ of protons and electrons, and an equation of state $P = 2nkT$, where $T=\frac{1}{2}(T_e+T_p)$ is the average of electron and proton temperatures. Mass conservation (\(4 \pi n u r^2 = \text{constant}\)), combined with the inviscid momentum conservation equation in a gravitational potential

\begin{equation}
nmu \frac{du}{dr} = - 2kT \frac{dn}{dr} - nm \frac{GM_\odot}{r^2}, \label{eq:momcon}
\end{equation}
yields
\begin{equation}
\frac{1}{u} \frac{du}{dr} \left( u^2 - \frac{2kT}{m} \right) 
= \frac{4kT}{mr} - \frac{GM_\odot}{r^2}. \label{eq:momcon_crit} 
\end{equation}
Here $u$ is the speed of radial expansion, $m$ is the sum of proton and electron masses, $k$ is the Boltzmann constant, $G$ is the gravitational constant, and $M_\odot$ is the solar mass. The right-hand side of Equation \eqref{eq:momcon_crit} vanishes 
at the \textit{critical radius} $r_c = GM_\odot m/4kT$. 
The left-hand side must also vanish here, 
for which we must have either a vanishing velocity derivative, or 
$u^2(r_c) \equiv u^2_c = 2kT/m$. 
The solutions of Equation \eqref{eq:momcon_crit} have the well-known `X', 
or \textit{saddle} type topology \citep[see e.g.,][]{hundhausen1972coronal};
the solution of physical interest is transonic, with a monotonically increasing velocity which is equal to the sound speed at the critical radius, i.e., at the 
sonic point.

As additional physical effects are added to a solar wind model, the mathematical structure of the equations changes, and with it the nature of the critical point \added{\citep[e.g.,][]{lamers1999book}}. For instance, including electrons in a two-fluid model would introduce two sound speeds and two possible critical points. As we will see in Section \ref{sec:results}, inclusion of the electron pressure in a two-fluid model shifts the location of the sonic point to a slightly greater heliocentric distance. Therefore, the ``singular'' aspect of a critical point is of limited physical relevance and it is questionable whether spacecraft data may be used to localize a definite critical point. \added{Observational and instrumental issues aside, the sharp transitions between regions of interest that emerge in the simplest models will almost certainly become more gradual transitions, or even ``fuzzy'' or erratic transitions, in the real solar wind that is influenced by three-dimensional (3D) effects, multifluid plasma physics, turbulence, etc.  In the following we will refer to these transitions as ``surfaces'' when it causes no confusion, but we remind the reader that in general we intend nonsingular and more gradual transitions \citep[see also][]{deforest2018ApJ}.}

\replaced{Nevertheless, From a physical perspective, these points (which become \textit{critical surfaces} in a three dimensional context) imply the existence of separate regions in the solar wind which are dominated by different physical effects.}{From a physical perspective, these critical points become critical surfaces in a 3D context and denote transitions between separate regions in the solar wind that are dominated by different physical effects.} For instance, counterpropagating Alfv\'enic fluctuations may effectively generate turbulence in the inner corona \citep{matthaeus1999ApJL523}, but above the Alfv\'en critical surface the population of inward propagating modes is diminished \citep{bruno2013LRSP}, and Alfv\'en wave collisions are no longer an efficient mode of turbulence production \citep{verdini2007apj}. The Alfv\'en surface also effects a separation of coronal regions having different angular flow properties; in the simplest picture, below this surface the torque produced by the magnetic field is sufficiently strong to transfer angular momentum and produce a corotation of the coronal wind with the sun, while above the critical surface the azimuthal velocity of the solar wind drops rapidly with distance \citep{weber1967ApJ148}.

In addition to the demarcation of different regions by critical surfaces, the general vicinity of the surfaces may be a site of interesting physics, such as enhancement in turbulent fluctuations \citep{lotova1985AA150}. These surfaces also signify the point beyond which MHD wave modes are unable to communicate upstream, because above the sonic (Alfv\'enic) critical surface the speed of propagation of information by sonic (Alfv\'en) modes is smaller than the speed of their advection downstream by the wind. Further, signatures of different coronal and solar phenomena may be evident in the location and morphology of critical surfaces, and may manifest in their temporal and spatial variability \citep{gral1996nature,lotova1997SoPh172}.

Recent observations by \cite{deforest2016ApJ828} and subsequent numerical investigations by \cite{chhiber2018apjl} provide additional \deleted{\textit{current}} motivation for the present study. Making use of highly processed \textit{STEREO} images from December 2008, \cite{deforest2016ApJ828} found a textural shift in the solar wind flow between heliocentric distances of 20 -- $80~R_\odot$. The images revealed that radially aligned, ``striated'' patterns gave way to more isotropic structures, \added{that DeForest et al.} termed ``\textit{flocculae}'', at distances of a few tens of solar radii. \cite{chhiber2018apjl} performed global solar wind MHD simulations, representing nominal large-scale solar wind conditions during December 2008, and superposed plasma-$\beta$ unity surfaces computed from these simulations on the \textit{STEREO} images. They found that the observed textural shift occurred near the first plasma-$\beta=$~1 surface. The emerging interpretation states that as the solar wind passes into the region where $\beta \equiv 8\pi P/B^2 \geq 1$, mechanical pressure may overcome the organizing influence of the magnetic field $B$, thus enabling the observed isotropic motions, which may be triggered by hydrodynamic shearing between wind streams \citep[e.g.,][]{roberts1992jgr}. A further point of interpretation, consistent with the one above, is that the \textit{flocculae} may be a manifestation of solar wind fluctuations interacting at the largest scales that are causally related through turbulence in the expanding solar wind \citep{chhiber2018apjl}.
The existence of such a maximum length scale of interaction is clear 
based on the finite amount of available propagation time, combined with
the assumption that the relevant correlations
must be produced by signals propagating at magnetohydrodynamic speeds. 
 
The Alfv\'en and $\beta=1$ surfaces may also be of significance to the phenomenon of preferential ion heating in the solar wind \citep[e.g.,][]{marsch2006kinetic}. Recently, \cite{kasper2017apj} found evidence for a zone, extending from just above the transition region ($\sim 0.3~R_\odot$) to a distance of tens of solar radii, where $\alpha$-particles are heated preferentially over protons. The outer boundary of this zone is likely associated with the Alfv\'en and $\beta=1$ surfaces. This point will be discussed further in Section \ref{sec:results}.
\section{Sampling of the three dimensional heliosphere by Parker Solar Probe}\label{sec:sampl}
The preceding section serves to 
emphasize the importance and relevance of critical surfaces. 
Yet, spacecraft missions hitherto have not been able to sample these in-situ (prior to \textit{PSP}, the closest heliocentric distance of approach was that of  \textit{Helios} at 0.29 au ($\sim62~\rs$)). \psp is 
set to change this by spending ``a total of 937 hours inside $20~\rs$, 440 hours inside $15~\rs$, and 14 hours inside $10~\rs$'' over its 7-year nominal mission \citep{fox2016SSR}. The spacecraft will most likely spend a very substantial amount of time under the first $\beta=1$ surface, which is inferred to lie between 20 and $60~\rs$ \citep{deforest2016ApJ828,chhiber2018apjl}.\footnote{The location of the Alfv\'en and first unit beta surfaces may dip below $10~\rs$ at the heliospheric current sheet (HCS). It must be noted that global models are likely to overestimate the spatial extent of the HCS due to their coarse resolution. This issue is discussed further in Section \ref{sec:results}.} According to observations and models \citep[e.g.,][]{mullan1990aa,lotova1997SoPh172,suzuki2005ApJ,cranmer2007ApJS,
verdini2010ApJ,pinto2011ApJ,oran2013ApJ,deforest2014ApJ787,
pinto2017ApJ,chhiber2018apjl,perri2018JPP}, the Alfv\'en surface lies between \(\sim\) \replaced{10}{2}\(\dash 30~\rs\) and \psp could spend a substantial time under this surface as well. The sonic surface may lie below the \textit{PSP}'s lowest perihelion at $9.86~\rs$, since coronal models often predict a location of $2\text{ -- }5~\rs$, although these predictions are applicable mainly to coronal hole regions \added{ \citep{kopp1976SoPh,mckenzie1995AA,habbal1995grl,
giordano2000apj,cranmer2007ApJS,verdini2010ApJ}}. At low latitudes the sonic point may lie as far as $20~\rs$ \citep{lotova1997SoPh172}. 
Since the periods in which the spacecraft 
will probe the regions within these surfaces will be of special
significance to the success of the \psp mission, it becomes a matter of some importance to estimate when these periods might occur. 

\begin{figure}
\gridline{\fig{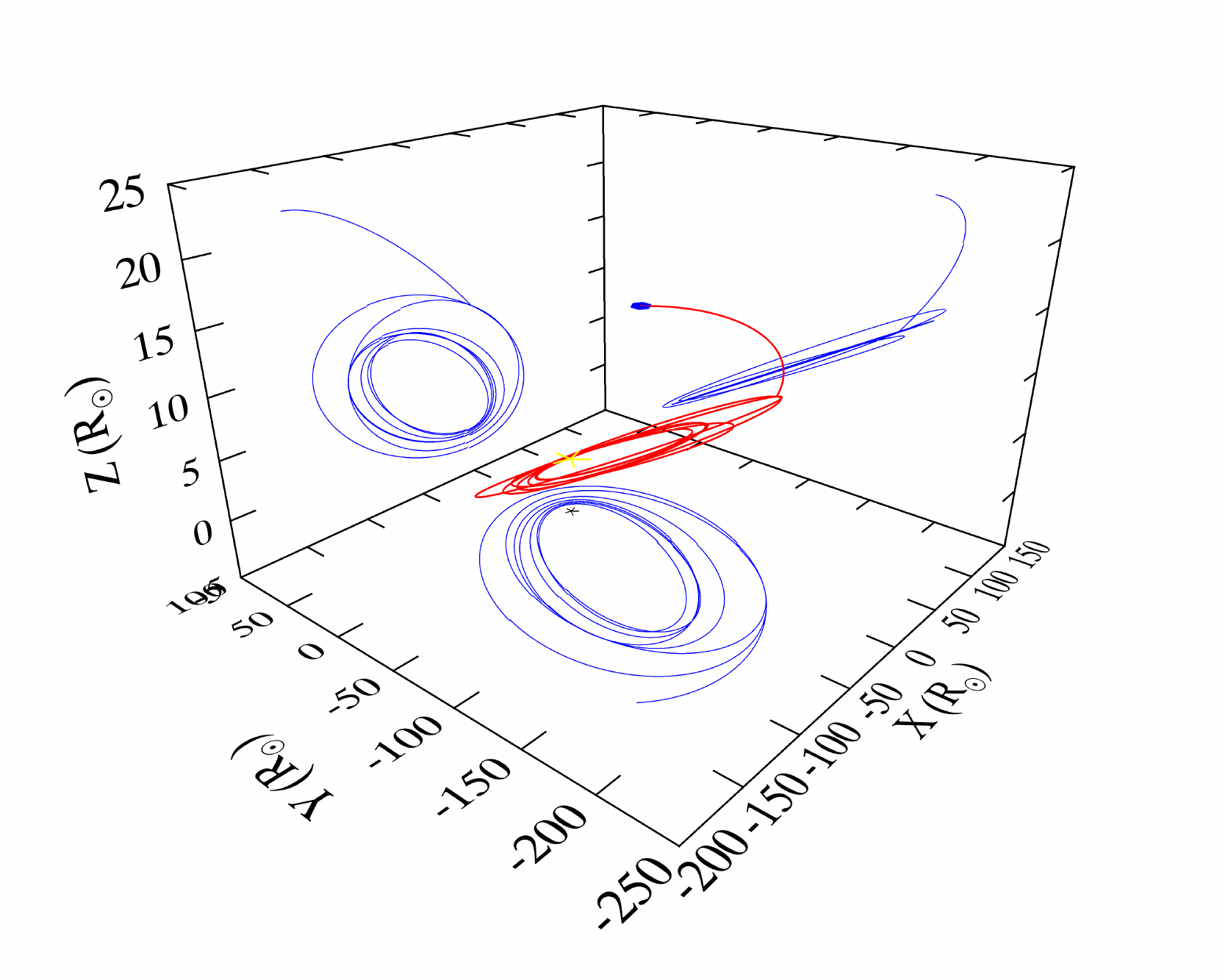}{.5\textwidth}{}}
 \caption{\psp trajectory in HCI coordinates (see text for details). The origin is the Solar center-of-mass and the $XY$-plane is the Solar equatorial plane. The red curves show the trajectory in 3D space and the blue curves are its projections onto the $XY, XZ,$ and $YZ$ planes. The `*' symbol and blue dot represent the positions of the Sun and Earth, respectively.}
\label{fig:overview}
\end{figure}

Figure \ref{fig:overview} shows a 3D perspective of the \psp trajectory. The spacecraft ephemeris was extracted from a \href{https://naif.jpl.nasa.gov/naif/index.html}{NASA SPICE kernel}, and the trajectory is presented here in the Heliocentric Inertial (HCI) coordinate system \citep[e.g.,][]{franz2002pss}. 
  Here the $XY$-plane is defined by the Sun's equator of epoch J2000; the $+Z$-axis is parallel to the Sun's rotation axis of epoch J2000, pointing toward the Sun's north pole; the $+X$-axis is the ascending node of the Solar equatorial plane on the ecliptic plane of J2000; and the origin of the coordinate system is the Sun's center of mass. The \psp trajectory in 3D space is shown in red, while the blue curves represent projections of the 3D trajectory onto the $XY, XZ$, and $YZ$ planes. The Earth (at time of launch) and the Sun are represented by the blue dot and the `*', respectively (not to scale). The trajectory shown includes all orbits in the 7-year nominal mission duration.

As \psp makes its high-resolution in-situ measurements, a knowledge of the large-scale environment within which these observations exist is of vital importance. In the next section we describe the solar wind model we have used to study the critical surfaces/regions and to make context 
predictions for the \psp trajectory.
\section{Solar Wind Model}\label{sec:model}

\deleted{A long-standing problem in heliospheric physics has been the identification of physical mechanisms that heat and accelerate the solar wind \citep{hundhausen1972coronal,leer1982ssr,meyervernet2007}, in particular the fast wind that emanates from coronal holes. The source of this additional energy presumably lies in the solar photosphere \citep{cranmer2005apjs}, but it must be transported across the chromospheric transition region and dissipated in the corona. Candidate mechanisms that enable this transport and dissipation include magnetic reconnection, wave and wave-particle interactions, and turbulence. Investigation of the finer details of these processes requires a kinetic description \citep{schekochihin2009ApJS182,servidio2015JPP,howes2017pop, yang2017pop}, but}

The large-scale features of the solar wind flow 
are widely regarded as well-represented in a fluid (MHD) description \citep{tu1995SSRv,goldstein1995araa,bruno2013LRSP,
matthaeus2015ptrs,makwana2015pop,parashar2015apj}.\footnote{One objection might be that magnetosonic modes may be heavily damped in kinetic theory \citep{barnes1979inbook};
an effect absent in MHD. However, compressive
modes may represent a small fraction of the energy in the weakly compressive 
interplanetary medium, and in any case the 
dissipation rate due to linear damping may be small compared to the 
cascade rate that leads to turbulent dissipation \citep{matthaeus2014apj}.} The MHD description is particularly indispensable for global simulation of the solar wind \added{\citep[e.g.,][]{gombosi2018LRSP}}, where the largest length scales in the system span at least a few solar radii ($1~\rs = 6.9 \times 10^5$~km). Kinetic effects come into play at the ion-inertial scale, which is roughly 90 km at 1 au \citep[e.g.,][]{schekochihin2009ApJS182} and becomes smaller closer to the sun. Current and foreseeable computational resources do not permit the resolution of this wide range of scales \citep[e.g.,][]{schmidt2015LRCA,miesch2015SSR194}. This makes MHD simulation our tool of choice for the current study that focuses on the global context of \psp observations. However, special provisions need to be made to 
preserve essential physical information contained in the smaller-scale 
\textit{fluctuations}, which are necessarily unresolved, even if the macroscopic 
features are well represented.
The large scales traversed by \psp orbits are illustrated strikingly in Figure \ref{fig:overview}, which serves to reinforce the appropriateness of this approach.

Fluid models of the solar wind have adopted various approaches to the problem of incorporating a source of heating and acceleration, including parametric heat deposition \citep[e.g.,][]{habbal1995grl,mckenzie1995AA,riley2015ApJ}, \added{a polytropic equation of state \citep[e.g.,][]{lee2009SoPh,gressl2014SoPh}}, WKB waves in a weakly inhomogeneous background \citep[e.g.,][]{jacques1978ApJ,usmanov2000global}, and MHD turbulence driven by Alfv\'en waves interacting with large-scale gradients \citep[e.g.,][]{matthaeus1999ApJL523,dmitruk2002ApJ575,suzuki2005ApJ,
verdini2010ApJ,vanderholst2014ApJ,yang2016SoPh}. We use an approach with a fully self-consistent and dynamical coupling of bulk solar wind flow with small-scale MHD turbulence -- bulk flow influences the turbulence, and in turn, turbulence dynamically feeds back into the bulk wind flow. In addition to turbulent heating and acceleration, the model incorporates two-fluid energy equations, heat conduction due to electrons, and proton-electron Coulomb collisions. We briefly describe \replaced{the bulk flow equations}{the model} below, 
and refer the reader to \cite{usmanov2018} for details, including those of the turbulence model and closure approximations.

Formally, the model is based on \deleted{the Reynolds-averaged Navier-Stokes approach, with} a Reynolds decomposition \citep[e.g.,][]{Monin1971book} applied to MHD. All physical fields, e.g., $\tilde{\mathbf{a}}$, are separated into a mean and a fluctuating component: \(\tilde{\mathbf{a}} = \mathbf{a}+\mathbf{a'}\), making use of an averaging operation where \(\mathbf{a} = \langle \tilde{\mathbf{a}} \rangle\). This ensemble average is associated with the large scales of motion, assumed to be deterministic. The quantity $\mathbf{a'}$ is a fluctuating component, here assumed to be of arbitrary amplitude and random in nature. By construction $\langle \mathbf{a'} \rangle = 0$. 

The model \deleted{as implemented here}
assumes that the solar wind is a fully ionized proton-electron plasma. The two species are described as fluids with separate energy equations and it is assumed that the bulk velocity is the same for the two species \citep{hartle1968ApJ151,hundhausen1972coronal,isenberg1986JGR,
marsch2006kinetic}. \deleted{To derive the mean-flow equations,} The velocity and magnetic fields are Reynolds-decomposed into mean and fluctuating components: $\tilde{\mathbf{v}} = \mathbf{v}+\mathbf{v'}$ and $\tilde{\mathbf{B}} = \mathbf{B}+\mathbf{B'}$, and the decomposed fields are substituted into the momentum and induction equations in the frame of reference corotating with the Sun. The ensemble averaging operator $\langle .\rangle$ is  applied, yielding large-scale, mean flow equations: a continuity equation, a momentum equation, an induction equation, and two pressure equations. The dependent variables are the mean velocity in the corotating frame $\vb$, the mean magnetic field $\bb$, the number density $N_S$ and pressure $P_S$ of solar wind (thermal) protons, and the pressure of electrons $P_E$. Pressures are assumed to be isotropic and we neglect density and pressure fluctuations \citep{usmanov2014three,usmanov2018}. The mass density $\rho = m_p N_S$ is defined in terms of the proton mass \(m_p\).

We use the classical Spitzer formula \citep{spitzer1965,hartle1968ApJ151} for the proton-electron Coulomb collision time scale, and the electron heat flux below $5\text{ -- }10~\rs$ is approximated by the classical collision dominated model of \cite{spitzer1953PhRv} \citep[see also][]{chhiber2016solar}, while above $5 \text{ -- } 10~\rs$ we adopt Hollweg's ``collisionless'' model \citep{hollweg1974JGR79,hollweg1976JGR}. Four turbulence quantities arise in the mean-flow equations: a source term $Q_\text{T}$ of energy deposition/extraction due to turbulent dissipation, the Reynolds stress $\Rb = \langle \rho \mathbf{v}'\mathbf{v}' - \mathbf{B}'\mathbf{B}'/4\pi \rangle$,
the magnetic pressure of the fluctuations \(\langle B'^2\rangle/8\pi\), 
and the mean turbulent electric field $\varepsilonb_m = \langle \mathbf{v}' \times \mathbf{B}' \rangle (4 \pi \rho)^{-1/2}$. These represent the coupling of the bulk flow to the small-scale fluctuations. Transport equations for the fluctuations are obtained by subtracting the mean-field equations from the full MHD equations. This yields a set of equations that describe the transport of three statistical descriptors for solar wind MHD fluctuations -- the turbulence energy, the correlation length of turbulent fluctuations, and the cross helicity -- which are coupled to the mean-field equations through terms involving $Q_\text{T}, \Rb,$ and $\varepsilonb_m$. To close the full set of equations, we employ an MHD analog of the familiar von K\'arm\'an--Howarth decay law \citep{karman1938prsl,wan2012JFM697,bandyopadhyay2018prx} for $Q_\text{T}$. Further details on the model, including those on numerical implementation, may be found in \cite{usmanov2014three} and \cite{usmanov2018}.

The simulations have been found to give reasonable agreement with many spacecraft observations of large-scale solar wind fields, turbulence parameters (energy, cross helicity, and  correlation scale), as well as the temperature, for varying heliocentric distance, and where feasible, varying helio-latititude \citep{breech2008turbulence,usmanov2011solar,usmanov2012three,
usmanov2014three,usmanov2016four,chhiber2018apjl,usmanov2018}. 
The model has been used to compute diffusion coefficients for energetic particles, again finding good agreement with spacecraft observations \citep{chhiber2017ApJS230}. Recent work (reviewed below) has combined our model's output with \textit{STEREO} images to enable a localization of the first $\beta=1$ surface \citep{chhiber2018apjl}.

The next section describes various runs of the simulation model 
performed for this work, and presents 
results 
relating to critical surfaces in the solar wind along 
with predictions along \psp orbits.
\section{Results}\label{sec:results}
The present work is based on analysis 
of 
two classes of simulation runs: 
(I) In the first case we employ a dipole magnetic field at the inner boundary, with the dipole tilted at angles of 0\degree, 5\degree, 10\degree, and 30\degree~(Runs I-A, I-B, I-C, and I-D, respectively) relative to the solar rotation axis. A 60\degree~run was also analyzed, but the results were found to be similar to the 30\degree~simulation. This simple configuration has both open (near the pole of the dipole) and closed (near its equator) magnetic field geometry, and allows for simulation of both coronal-hole-like and streamer-like flows. This gives us a representation of the ambient, large-scale bimodal solar wind flow during periods of low-to-medium solar activity \citep{mccomas2003grl,usmanov2003tilted,owens2013lrsp}.
(II) In the second case the MHD code is driven by a magnetic field at the base obtained from July 1989, July 1994, and December 2008 magnetogram data (Runs II-A, II-B, and II-C, respectively) published by the Wilcox Solar Observatory. Note that the 
magnetogram runs use a slightly older numerical model with a simpler WKB-wave based treatment of the coronal region  \citep[\(1 \dash 45~\rs\); see][]{usmanov2000global,usmanov2003tilted,usmanov2014three}, \added{since the new coronal model \citep{usmanov2018} requires further testing with boundary conditions based on solar-maximum magnetograms.}

The simulation domain extends from the coronal base at $1~\rs$ to 3 au. The following input parameters are specified at the coronal base: the
driving amplitude of Alfv\'en waves ($\sim 30$ km~s$^{-1}$), the
density ($\sim 1 \times 10^8$ particles cm$^{-3}$) and temperature
($\sim 1.8 \times 10^6$~K). The magnetic field magnitude is assigned either using a source magnetic dipole on the Sun's poles (with
strength 12~G to match values observed by Ulysses) or from solar magnetograms. Runs I-A to I-D use an adiabatic index  \(\gamma = 1.67\) throughout the simulation domain, while Runs II-A to II-C use \(\gamma=1.02\) in the WKB-based coronal region and \(\gamma=1.67\) above \(45~\rs\). For further numerical details see \cite{usmanov2014three} and \cite{usmanov2018}.
%
%
\subsection{Surfaces in the Meridional Plane}
The significance of the 
sonic and Alfv\'en critical surfaces, as well as the first $\beta=1$ surface, 
was discussed in Section \ref{sec:back}. Operationally the Alfv\'en critical surface is defined by the set of points, scanning outward, at which the solar wind speed first exceeds the Alfv\'en speed \(V_A = B/\sqrt{4 \pi \rho}\). Similarly, the sonic surface is defined by the set of points, scanning outwards from the sun, at which the total solar wind speed becomes larger than the sound speed $c_s = \sqrt{\gamma P_p/\rho}$. 
Here $\gamma$ is the polytropic index and $P_p$ is the proton pressure. Another definition of the sound speed is $c'_s = \sqrt{\gamma P/\rho}$,
where $P = P_p + P_e$ includes the electron pressure $P_e$. We show the sonic surfaces computed using both these definitions to stress that the inclusion of various physical effects may change the location of the surface, and it is perhaps more appropriate to envision a transonic \textit{region} \citep{lotova1997SoPh172} rather than a highly localized surface. Nevertheless, at the fluid level of description $P$ may be considered the more appropriate measure of pressure. 

The plasma beta is also defined in two ways; in terms of the proton beta, \(\beta_p = 8\pi P_p/B^2\), and in terms of the total electron plus proton beta, \(\beta_{p+e} = 8\pi (P_p + P_e)/B^2\). The first $\beta=1$ surface is  identified as the set of points, scanning outward, at which $\beta=1$ is first encountered. This is done in the analysis separately for proton beta and for total beta. 

\begin{figure}
\gridline{\fig{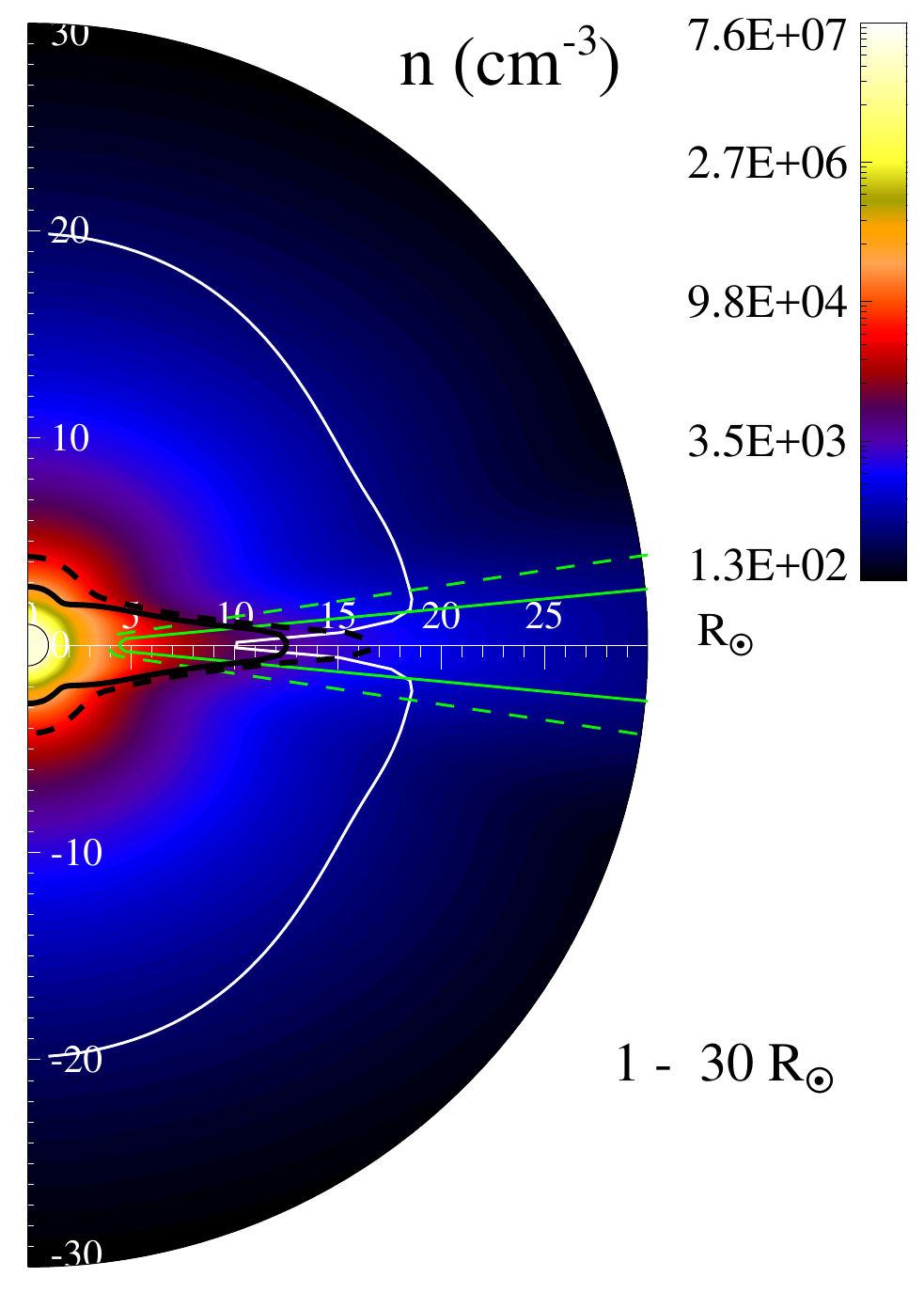}{0.25\textwidth}{(a)}
          \fig{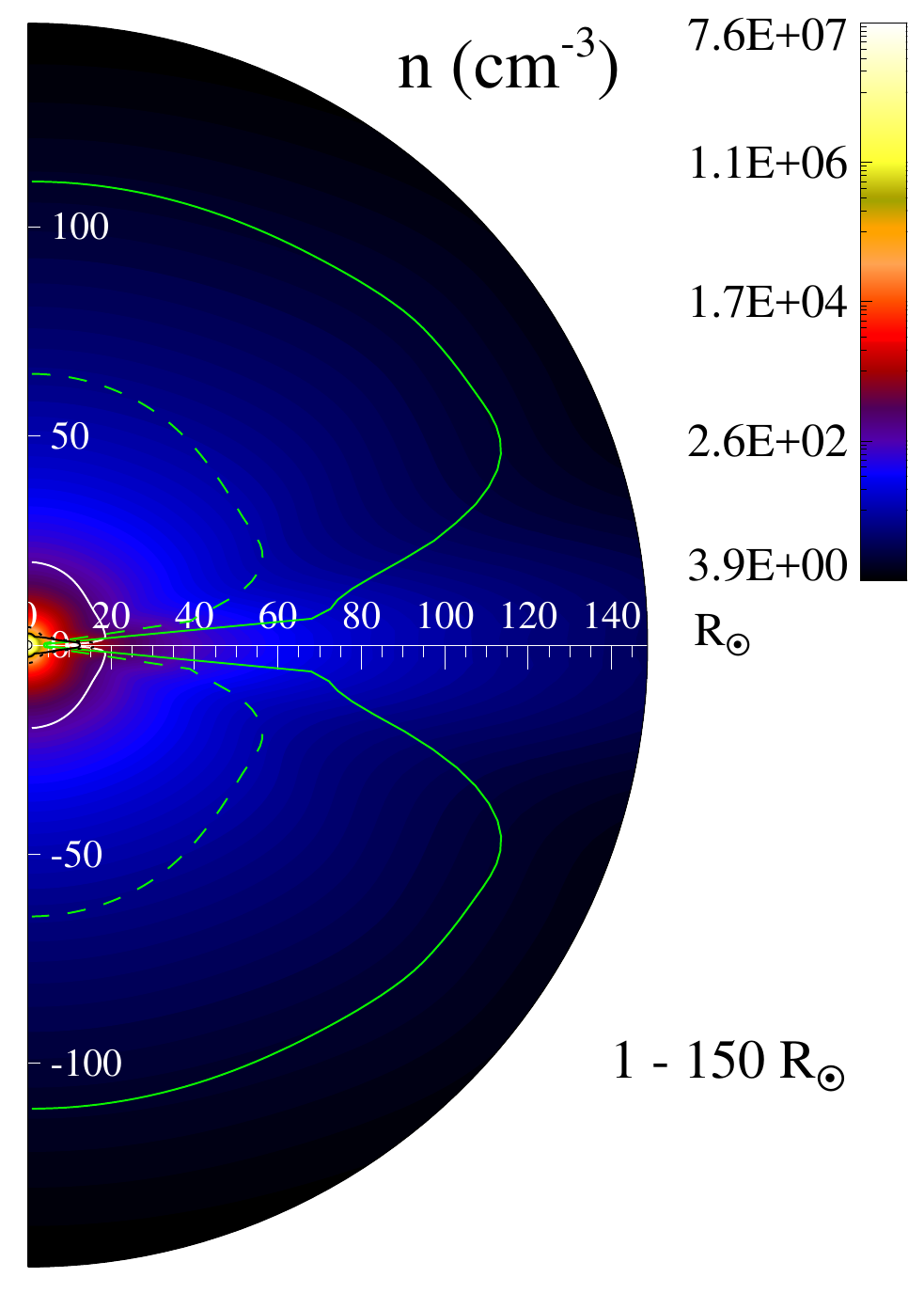}{0.25\textwidth}{(b)}
          \fig{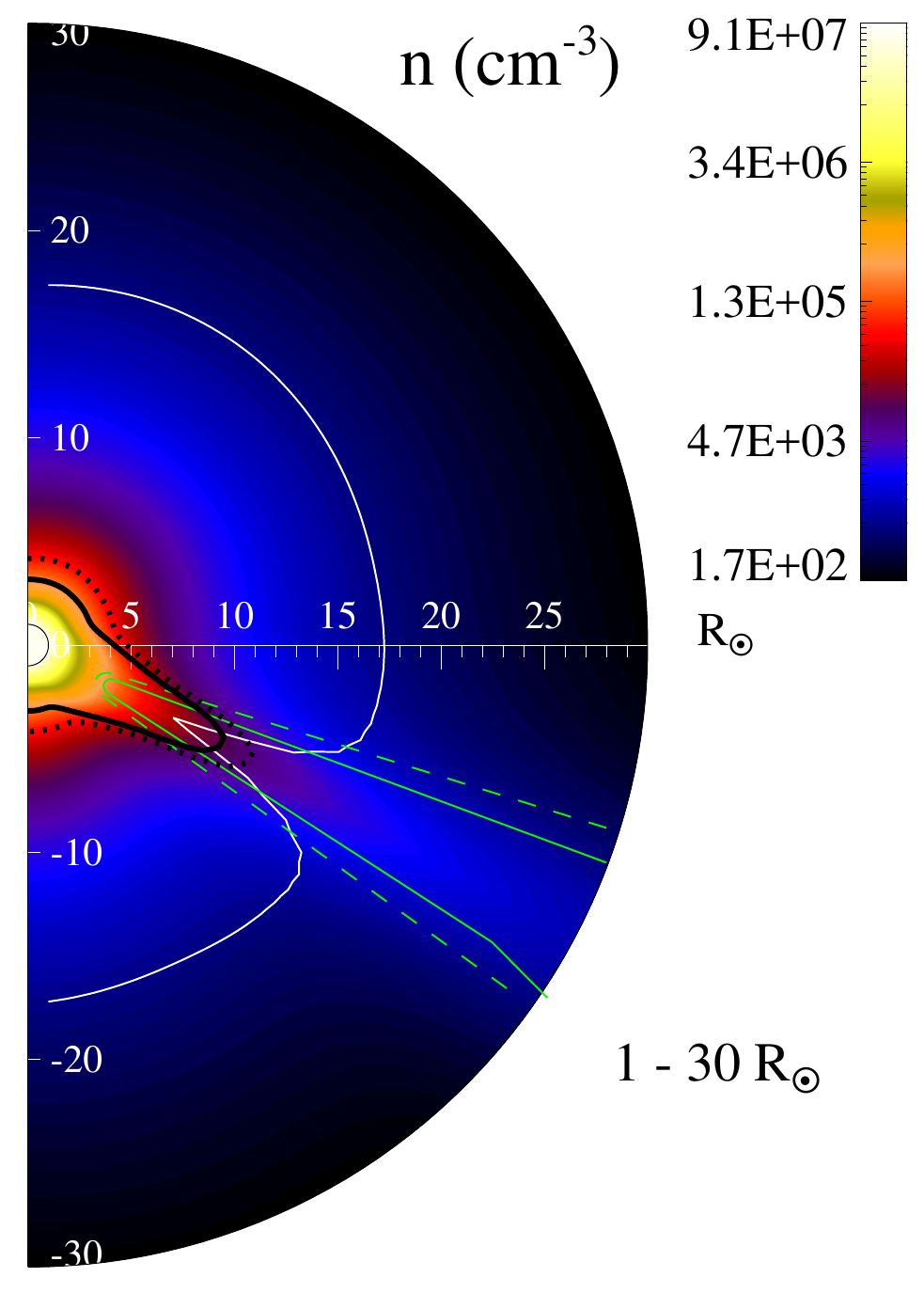}{0.25\textwidth}{(c)}
          \fig{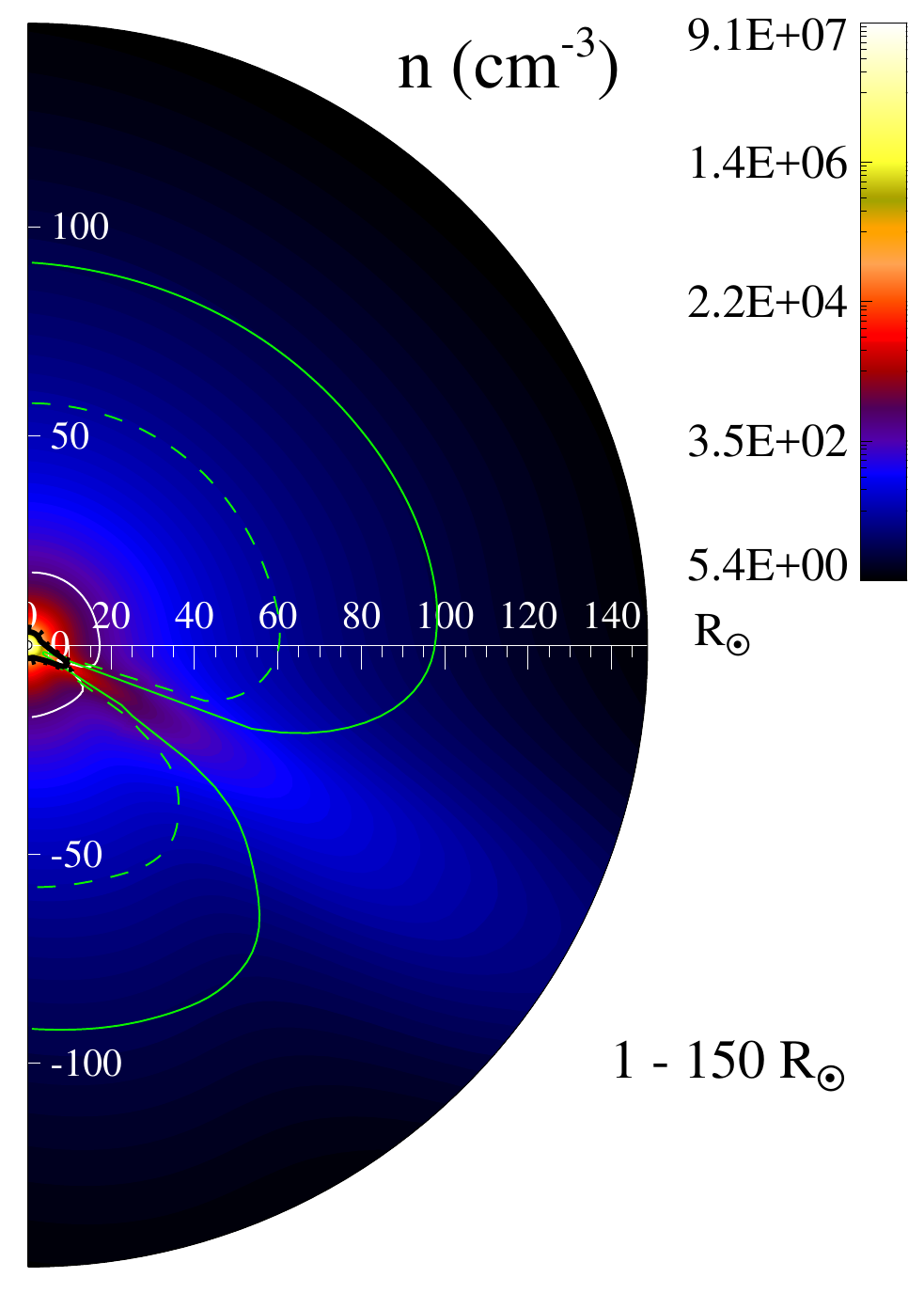}{0.25\textwidth}{(d)}
          }          
\caption{Meridional planes from untilted dipole Run I-A ((a), (b)) and 30\degree~tilted dipole Run I-D ((c), (d)). 
Panels (a) and (c) show heliocentric distances from \(1 \dash 30~\rs\) while panels (b) and (d) show \(1 \dash 150~\rs\).  
The black curves show the sonic surface (solid line using $c_s$ with just proton pressure and 
dashed line using $c'_s$ which includes proton and electron pressures; see text), 
the white curve 
shows the Alfv\'en surface, and the green curves show the first unity $\beta$ 
surface (solid line shows $\beta_p=1$ and dashed line shows $\beta_{p+e}=1$).}
\label{fig:merid}
\end{figure}

Figure \ref{fig:merid} depicts the projection of these surfaces 
onto an arbitrarily selected meridional 
plane at 37\degree~heliolongitude for Runs I-A and I-D. Unless specified otherwise, simulation data are 
plotted in the Heliographic Coordinate system \citep[HGC,][]{franz2002pss}. 
Heliographic latitude is measured from the solar equator positive 
towards North, Heliographic longitude is defined in the direction 
of planetary motion, with the $XY$-plane defined by the solar equator.

The surfaces show a laminar appearance, and display a very organized ordering. 
The two configurations depicted are very similar, with no asymmetry in the zero-tilt case, 
and only minor asymmetries seen in the north-south direction.
For all latitudes well-separated from the heliospheric current sheet (HCS), 
the $\beta=1$ surface is the most distant, with the Alfv\'en surface
contained well within it, and the sonic surface(s) lower still, in the range \(3 \dash 5~\rs\). 
The most dramatic feature is the rearrangement of the surfaces 
near the heliospheric current sheet region \added{\citep[consistent with previous work that examines the properties of these surfaces, e.g.,][]{pneuman1971SoPh,keppens2000ApJ,usmanov2000global,
pinto2011ApJ,oran2013ApJ}}, an effect that can completely  
reverse the surfaces to an opposite ordering. In fact, one can find a 
substantial region in which 
the $\beta=1$ surface lies at lower altitudes than the Alfv\'en surface.
There are also regions, much smaller in these particular cases, in which the sonic surface is found at altitudes above the Alfv\'en surface. 
In those small regions, the solar wind would have the somewhat anomalous character 
of being super-Alfv\'enic but subsonic. Alfv\'en wave pressure in such regions
may be able to increase the mass flux of the resulting wind at higher radial distances \citep[see][]{leer1982ssr}.

Before proceeding
with further analysis, we want to emphasize that
there are unavoidable limitations 
in using these simulations. 
One obvious comment is that our MHD solutions are based on simplified data that do not represent the actual boundary conditions 
corresponding to the solar wind 
during the \psp passage.  
More specifically, we emphasize that the 
discrete
spatial resolution 
of the MHD model limits 
the thinning of the HCS.
Therefore both the HCS and the much wider plasma sheet surrounding it 
are expected to be broader in the simulation than in the actual solar wind \citep{winterhalter1994JGR}.
A rough estimation based on published data suggests that
the real HCS may be a factor  of \(\sim 5\) thinner than what we are able to resolve here. 
Nevertheless, within the 
resolution parameters of the code, the physics of the simulation is 
deemed to be accurate, so that, for example, the 
inversion of critical surfaces is expected 
to occur, albeit over a thinner 
region, 
in the solar minimum conditions seen in some \psp orbits.\footnote{It would be of interest to compare the present MHD-based results with analyses based on flux-tube solar wind models in which the HCS remains thin \citep[e.g.,][]{pinto2017ApJ}. Such a comparison is outside the scope of the present paper.}
\subsection{Remote Sensing Context}
\begin{figure}
\centering
\includegraphics[scale=.2]{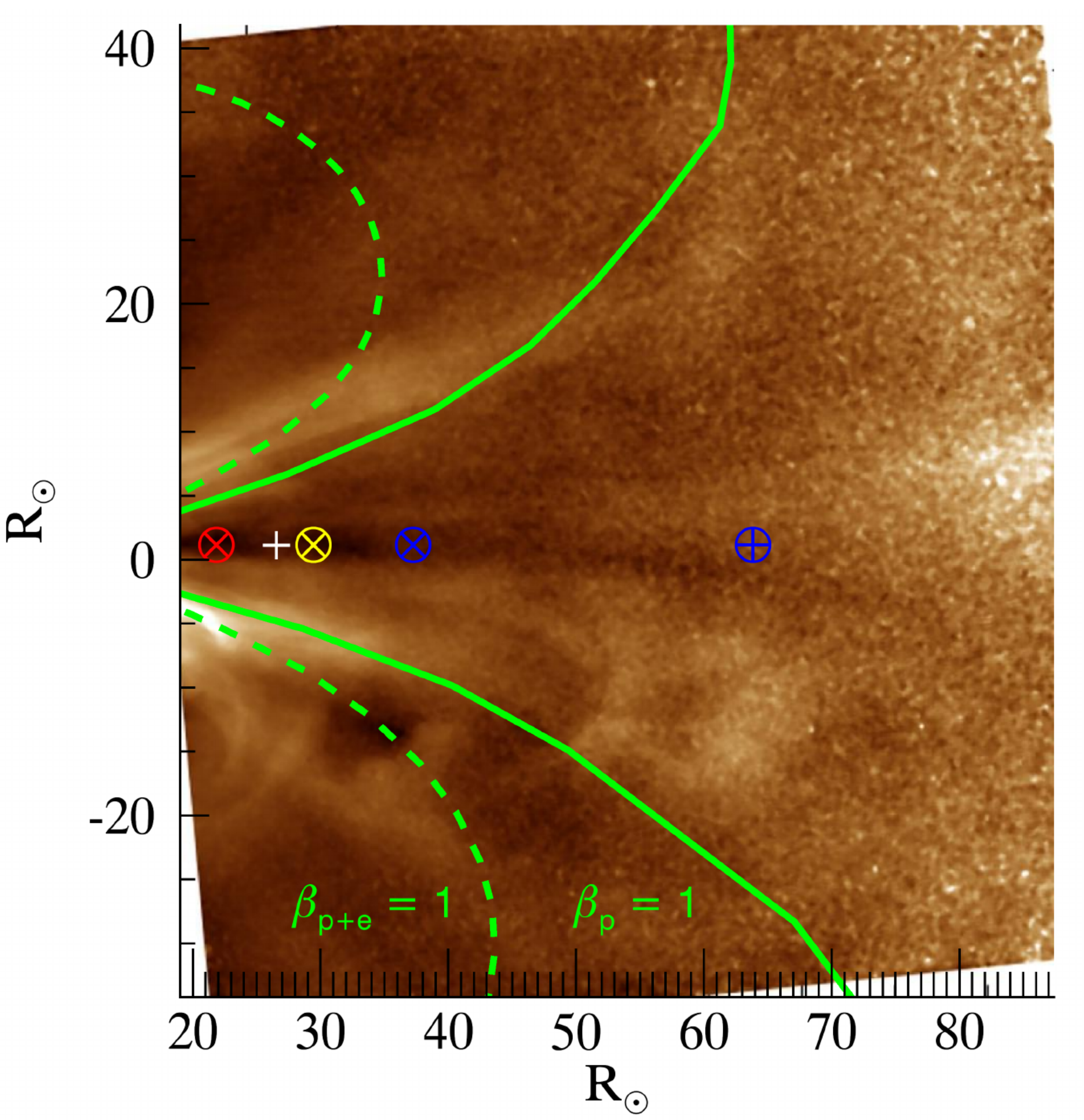}
\caption{Green curves show the first unity beta surfaces (solid line for $\beta_p=1$; dashed line for $\beta_{p+e}=1$) computed from the model (Run II-C) superimposed on a \textit{STEREO} image from \cite{deforest2016ApJ828}. White `+' shows location of enhanced turbulence inferred by \citet{lotova1985AA150} (see Figure \ref{fig:scintillation}); \textit{Helios} perihelion is shown as `{\color{blue}$\oplus$}'; the first three perihelia of the \psp are shown as `$\otimes$'.} 
\label{fig:stereo}
\end{figure}

We recall briefly the
novel use of \textit{STEREO} Heliospheric Imaging (HI)
data by \citet{deforest2016ApJ828}, which  
examined a series of images of the inner solar wind
and argued, based on physical grounds, 
that the observed striation-flocculation
transition
occurred in the neighborhood of the first plasma-\(\beta =1\)
surface.  
\cite{chhiber2018apjl}
employed MHD simulations, similar to those analyzed here,
to provide confirming evidence of this interpretation. Figure \ref{fig:stereo} revisits this analysis, showing that the region in which the 
\textit{striae} gives way to \textit{flocculae} 
is commensurate 
with the region in the simulation in which the first $\beta=1$
surface is encountered, as the wind transitions 
from magnetic control to hydrodynamic control. 

Recently, \cite{kasper2017apj} found evidence for a zone, extending from just above the transition region ($\sim 0.3~R_\odot$) to a distance of tens of solar radii, where $\alpha$-particles are heated preferentially over protons. The lower boundary of this zone would likely be at the chromospheric transition region, where the plasma collisionality changes from high to weak, thus permitting nonthermal physics to produce observed temperature anisotropies \citep[e.g.,][]{marsch2006kinetic}. It is conceivable that this zone of preferential heating ends at the first beta unity surface, since kinetic temperature anisotropies are generally associated with \(\beta \lesssim 1\) \citep[e.g.,][]{matteini2012SSR172}. This zone should be detected by the \psp when it reaches below the first beta unity surface.

\begin{figure}
\gridline{\fig{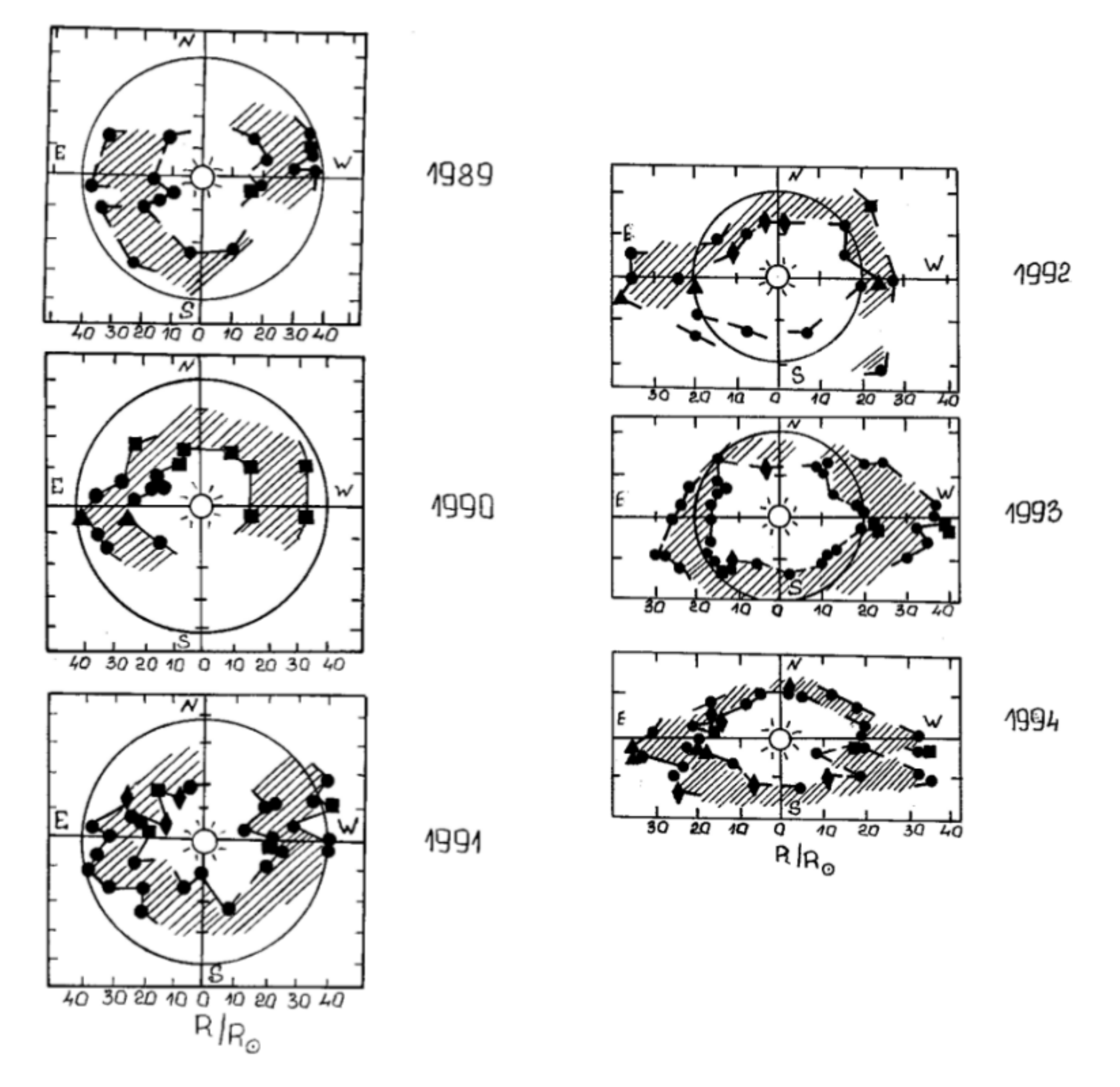}{0.5\textwidth}{(a)}
          }
\gridline{\fig{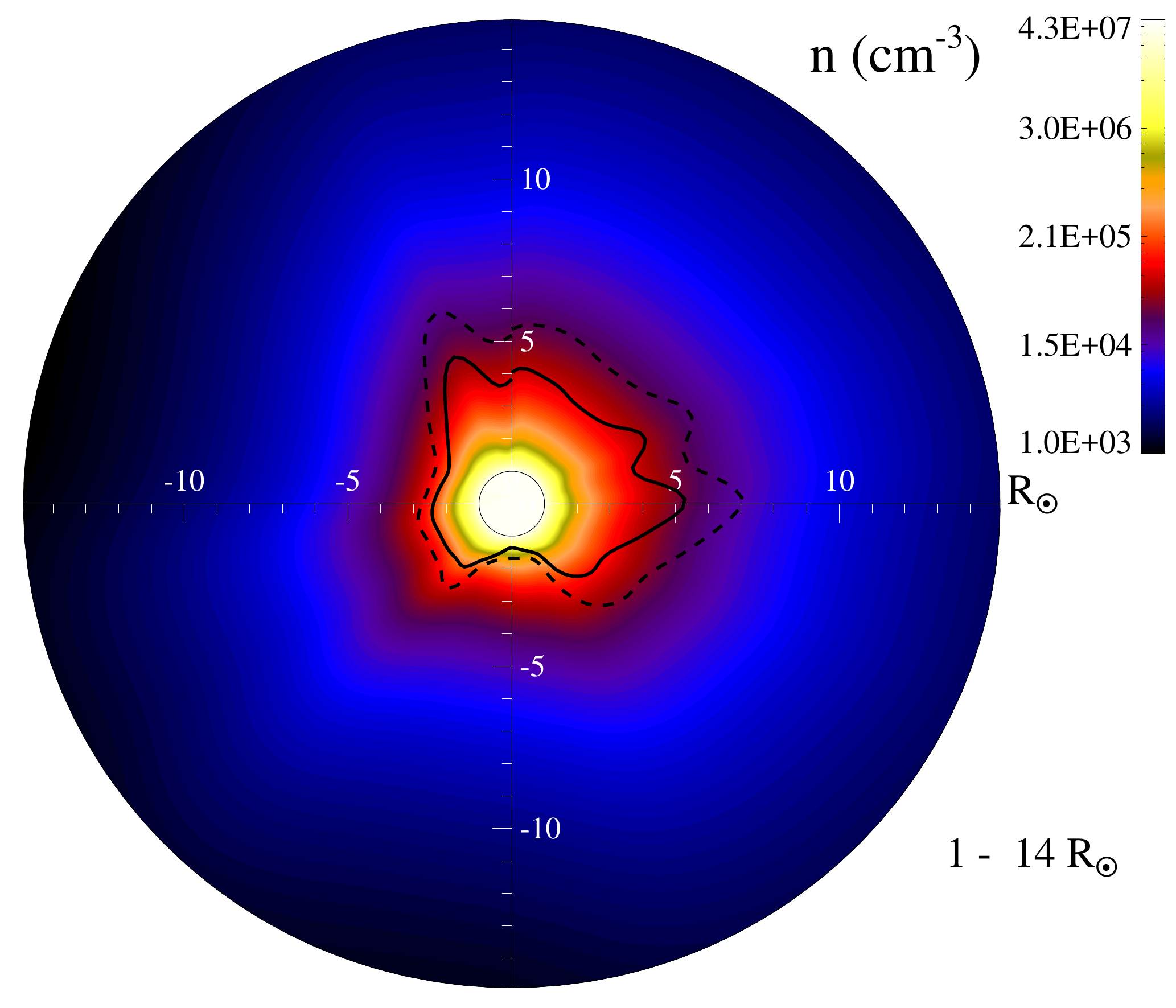}{0.4\textwidth}{(b)}
          \fig{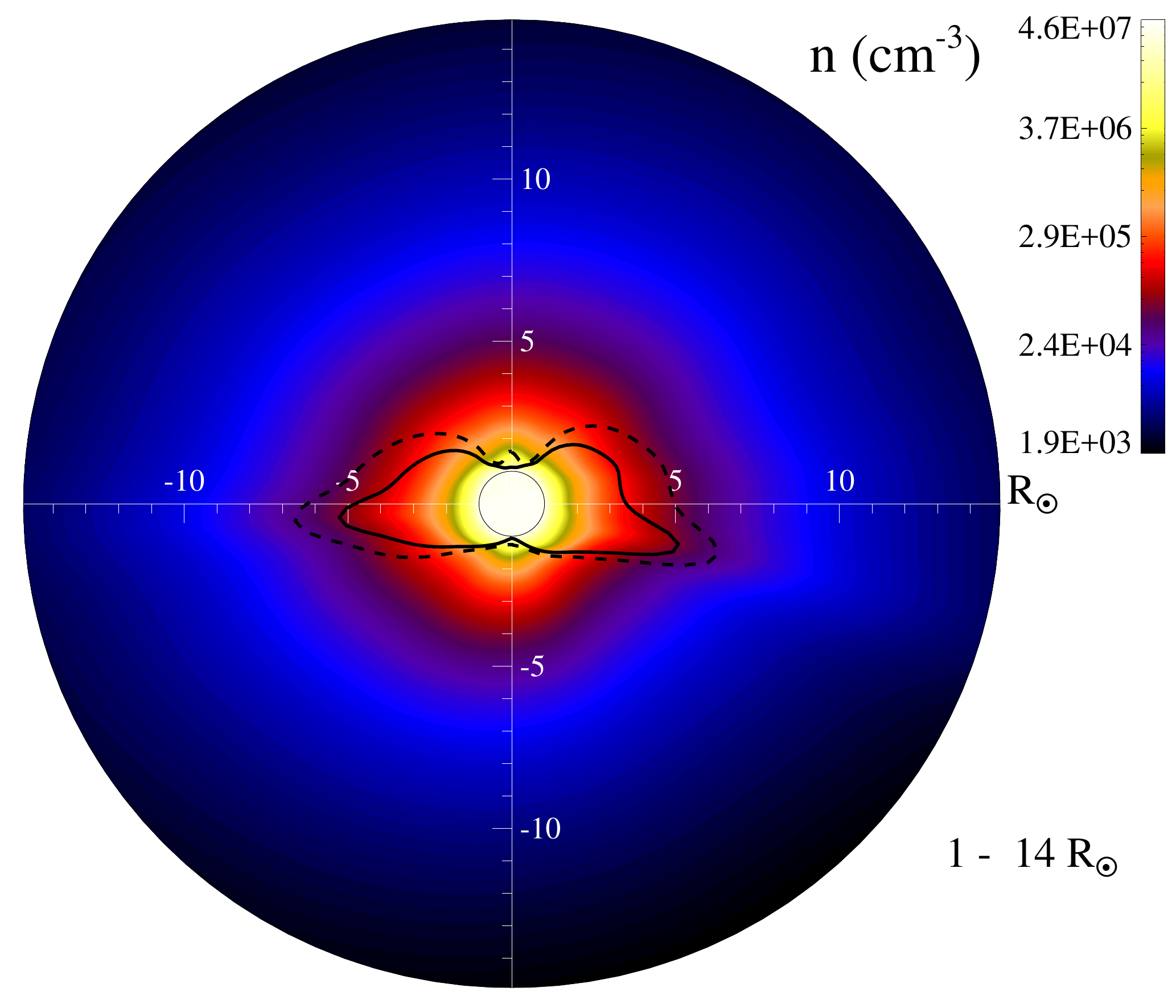}{0.4\textwidth}{(c)}
          }          
\caption{(a) Transonic regions from \cite{lotova1997SoPh172}, showing the transition from spherically symmetric but jagged morphology at solar maximum (1989), to oblateness at the poles during solar minimum (1994). Sonic surfaces (solid line using $c_s$ with just proton pressure and dashed line using $c'_s$ which includes proton and electron pressures; see text) from Runs II-A and II-B, using solar maximum (July 1989) and solar minimum (July 1994) magnetograms, respectively. Contours of proton density are shown in the background. The transition from solar maximum (b) to solar minimum (c) is qualitatively consistent with the one seen in Figure \ref{fig:lotova1997}(a).} 
\label{fig:lotova1997}
\end{figure}

The location of the sonic critical surface as a function of latitude
was estimated from scintillation data by \citet{lotova1997SoPh172}. Figure~\ref{fig:lotova1997} shows the Lotova et al. results and compares them with sonic critical surfaces 
obtained from two MHD simulations -- a solar minimum magnetogram
and a solar maximum magnetogram.
We note a reasonable qualitative similarity, especially regarding the
oblateness at the poles during solar minimum and the spherical but jagged shape during solar maximum. During solar minimum, there exists a clear demarcation between slow wind streams at equatorial latitudes and fast wind in polar regions. As a result, the wind becomes supersonic at larger distances from the Sun at low latitudes, while the sonic surface at the poles lies at lower heights. These results \replaced{suggest}{support the idea} that variations in the morphology of the critical surfaces can be used to infer the state of solar activity \added{\citep[e.g.,][]{keppens2000ApJ,pinto2011ApJ,pinto2017ApJ}}.

\begin{figure}
\gridline{\fig{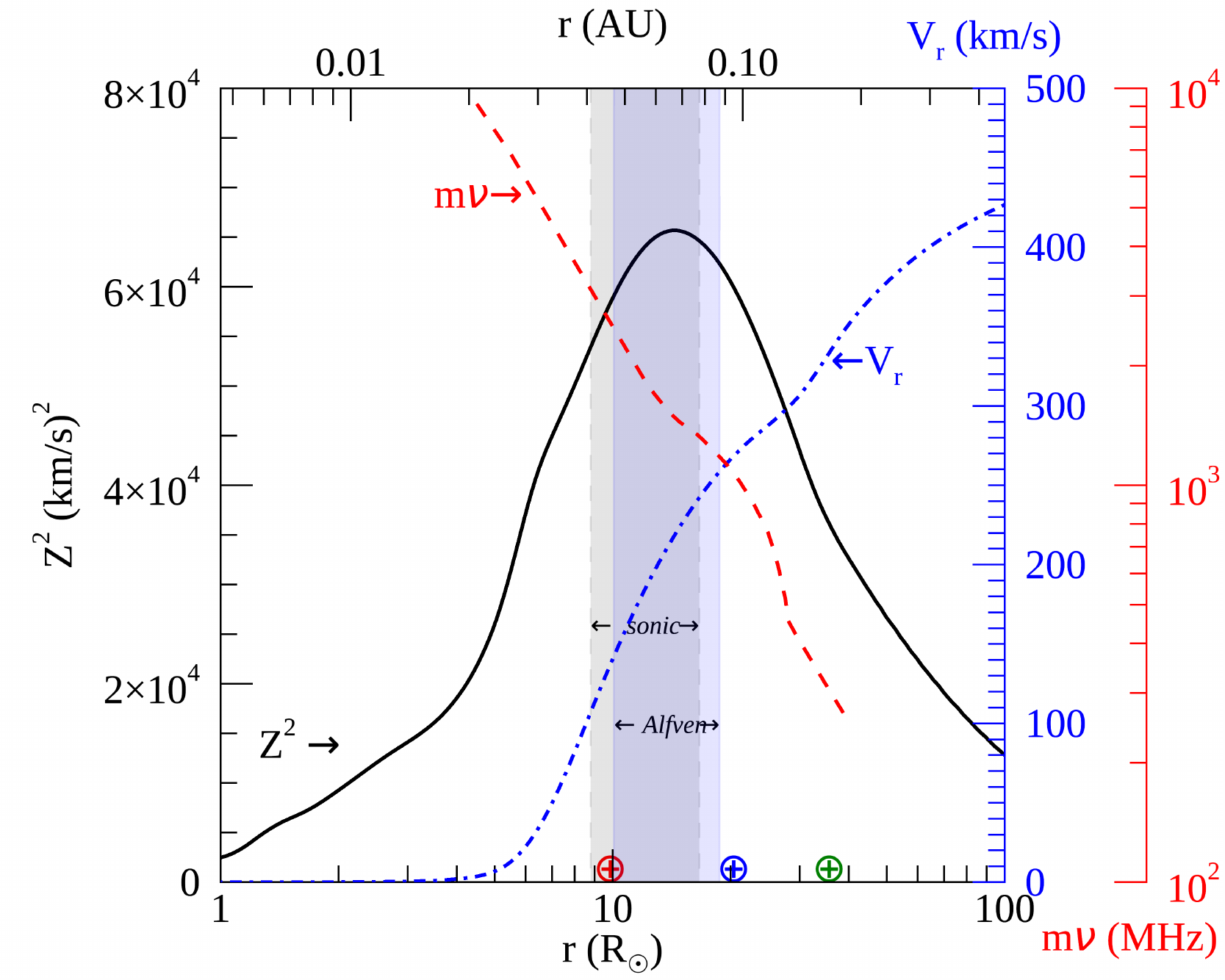}{0.5\textwidth}{(a)}
          \fig{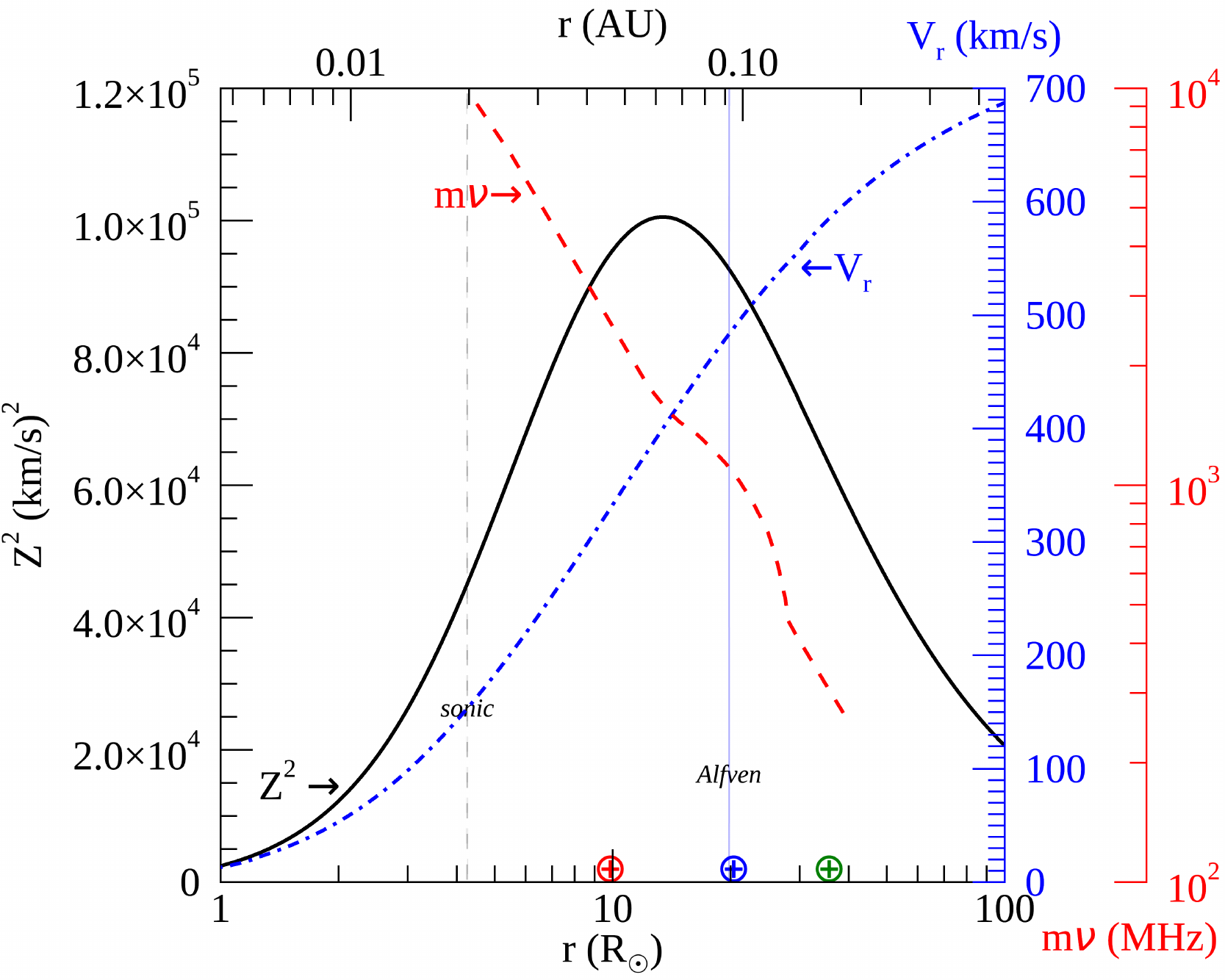}{0.5\textwidth}{(b)}
          }          
\caption{Enhanced scintillation ($m\nu$) region from the observations of \cite{lotova1985AA150}, seen as a bump at \(\sim 20~\rs\) in the dashed red curve. Radial solar wind speed \(V_r\) (dash-dotted blue curve) and turbulence energy density (per unit mass) \(Z^2\) (solid black curve) are shown at (a) an ecliptic heliolatitude of 6.75\degree~and (b) a polar heliolatitude of 82\degree. Panel (a) shows shaded bands representing the locations of the Alfv\'en (pale blue band) and sonic (grey band with dashed outline) surfaces in the ecliptic region of the simulation (between heliolatitudes 6.75\degree~and \(- 6.75\degree\)). Panel (b) shows vertical lines representing locations of the Alfv\'en surface (pale blue solid) and the sonic surface (grey dashed) at 82\degree~heliolatitude. All simulation results shown here are from Run I-A. The first, third, and final perihelia of the \psp are represented as $\oplus$ symbols, {at heliocentric distances of 35.66, 20.35, and 9.86\(~\rs\), respectively \citep{fox2016SSR}.}} 
\label{fig:scintillation}
\end{figure}

Another look at the properties of the solar wind in the critical region
is provided by the 
scintillation intensity data of \cite{lotova1985AA150}, reproduced in Figure \ref{fig:scintillation}. 
For comparison we show the radial profiles of two parameters obtained \deleted{with} from an \added{(axisymmetric)} simulation with an untilted dipole (Run I-A), \added{in the ecliptic (Figure \ref{fig:scintillation}(a)) and polar (Figure \ref{fig:scintillation}(b)) regions.} The parameters shown are the radial solar wind speed \(V_r\) and the turbulence energy density (per unit mass) \(Z^2\) \added{at 6.75\degree~heliolatitude, representative of the ecliptic region (Figure \ref{fig:scintillation}(a)), and at 82\degree~heliolatitude, representative of the polar region (Figure \ref{fig:scintillation}(b)).} The scintillation profile (measured through $m\nu$, where \(m\) is a scintillation index and \(\nu\) is the frequency of observation; see \cite{lotova1985AA150}) shows a feature in the range of 
\(15\dash 30~\rs\) that is interpreted as a region of enhanced turbulence, giving rise to enhanced radio scattering from density irregularities. 
Shaded regions in the Figure \ref{fig:scintillation}(a) indicate the range of radii at which 
the Alfv\'en and sonic surfaces are found in the ecliptic region in the simulation \added{(between heliolatitudes 6.75\degree~and \(- 6.75\degree\)), while the vertical lines in Figure \ref{fig:scintillation}(b) represent the locations of these surfaces at 82\degree~heliolatitude. The Figure also shows \psp perihelia for several orbits.} 
We note that the scintillation feature lies very close to the position of the maximum turbulence energy per unit mass $Z^2$
from the simulation, and is also close to the locations of the sonic and Alfv\'enic critical surfaces in the simulation. 
This enhancement in turbulence may be caused by the interactions of counter-propagating Alfv\'en waves \citep{matthaeus1999ApJL523}. The acceleration of the wind also begins in this region, \added{with larger speeds and turbulence energies seen at polar latitudes.}
%
%
\subsection{What PSP will see: Dipole-based Simulations}
Using the \psp trajectory and a coordinate
transformation to link it to the global MHD solution,
one may graphically illustrate the 
relationship between 
the \psp orbit and the 
simulated heliospheric structure. 
Superposing the 
orbits on the simulation results should not be construed as a prediction,
since the boundary data, even if compatible 
with projected future conditions, is necessarily imprecise. 
However this exercise does present a possible
context for the \psp mission.
\replaced{Portraying this relationship is not trivial, because the critical surfaces rotate with Sun, while the \psp orbit traces a curve in three-space that does not precisely lie in a single plane in any inertial frame (see Fig. \ref{fig:overview}).}{Here we evaluate the MHD solution along the \psp trajectory, taking solar rotation into account.}

\begin{figure}
\gridline{\fig{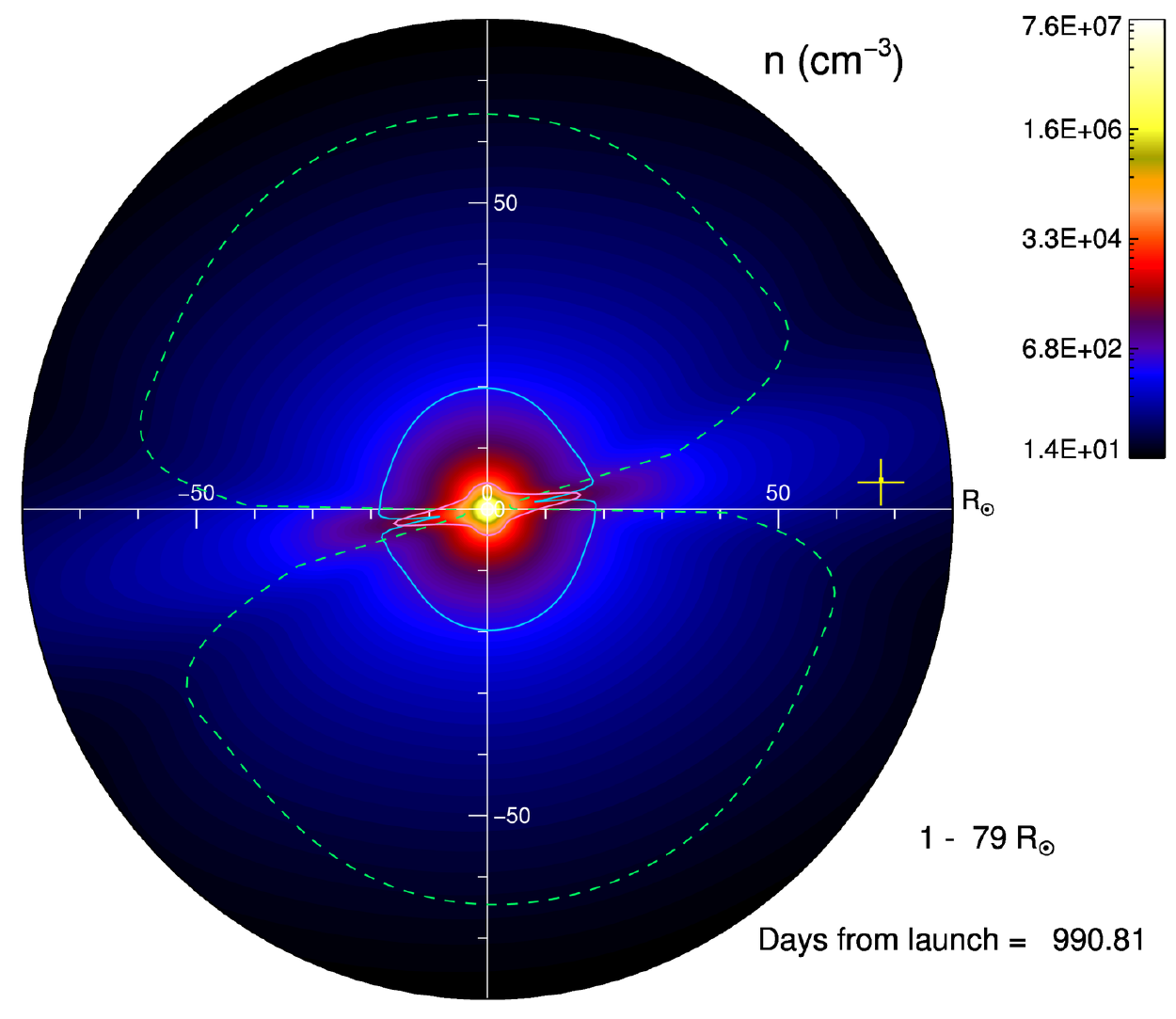}{0.45\textwidth}{(a)}
          \fig{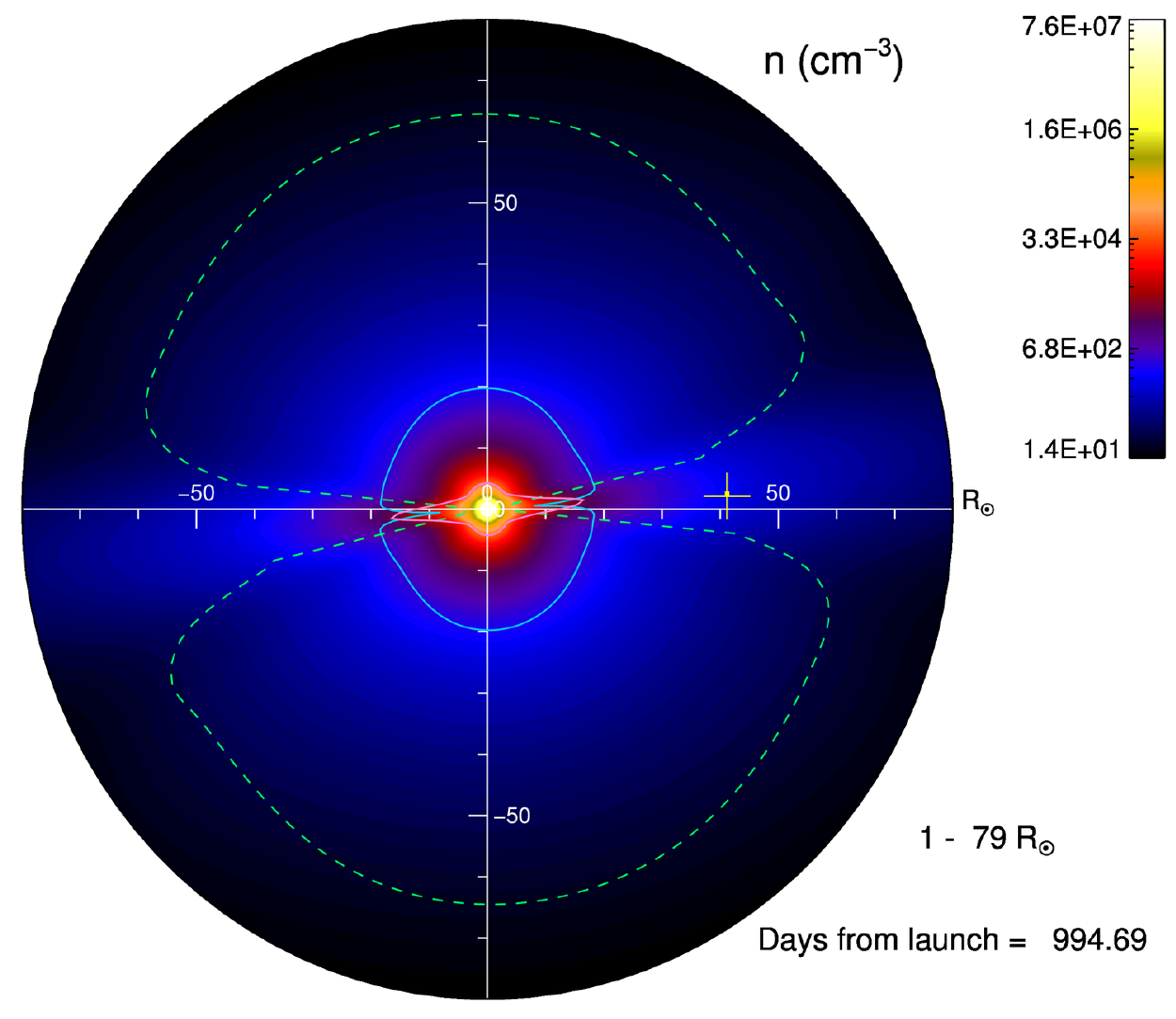}{0.45\textwidth}{(b)}
          }
\gridline{\fig{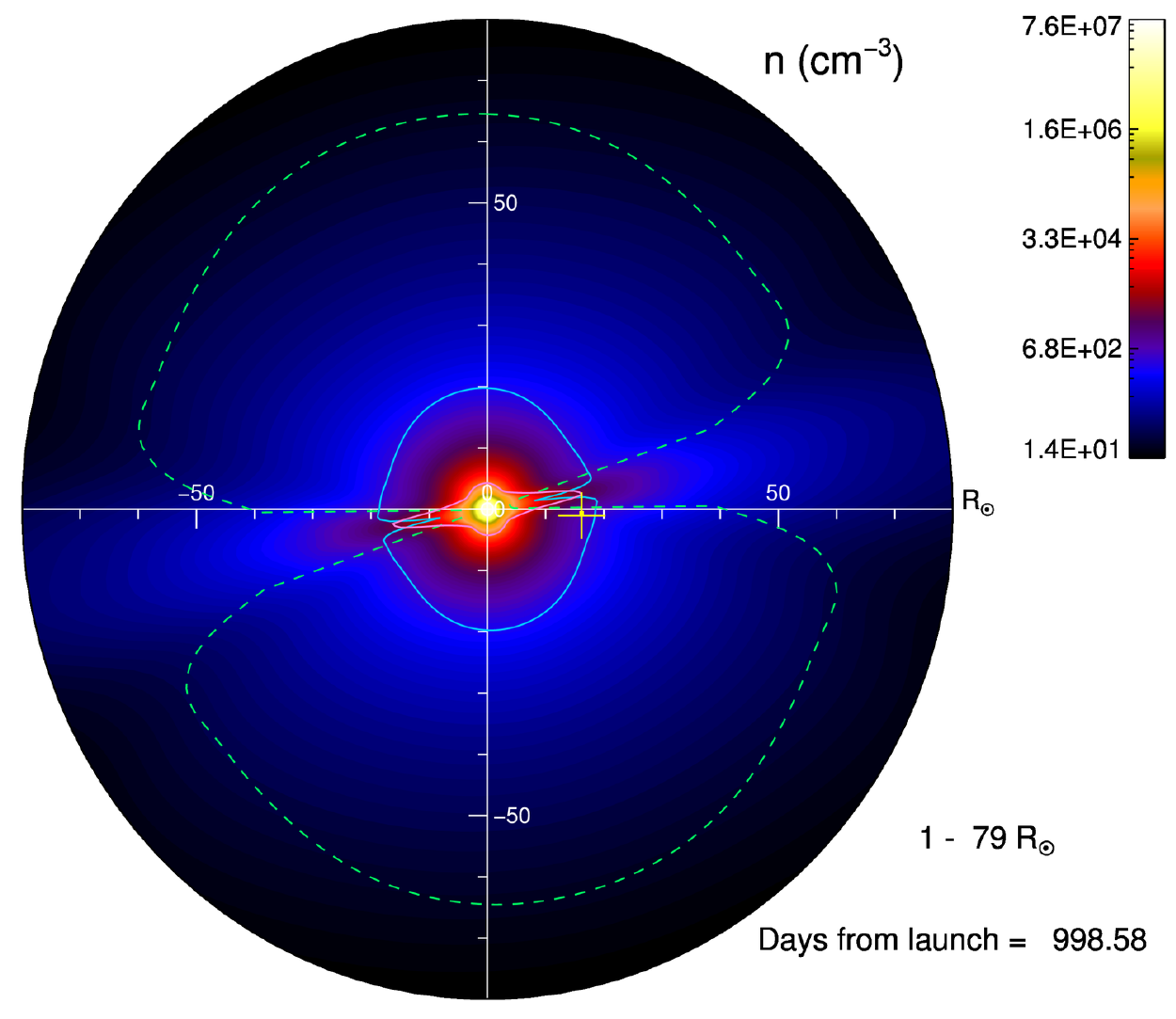}{0.45\textwidth}{(c)}
          \fig{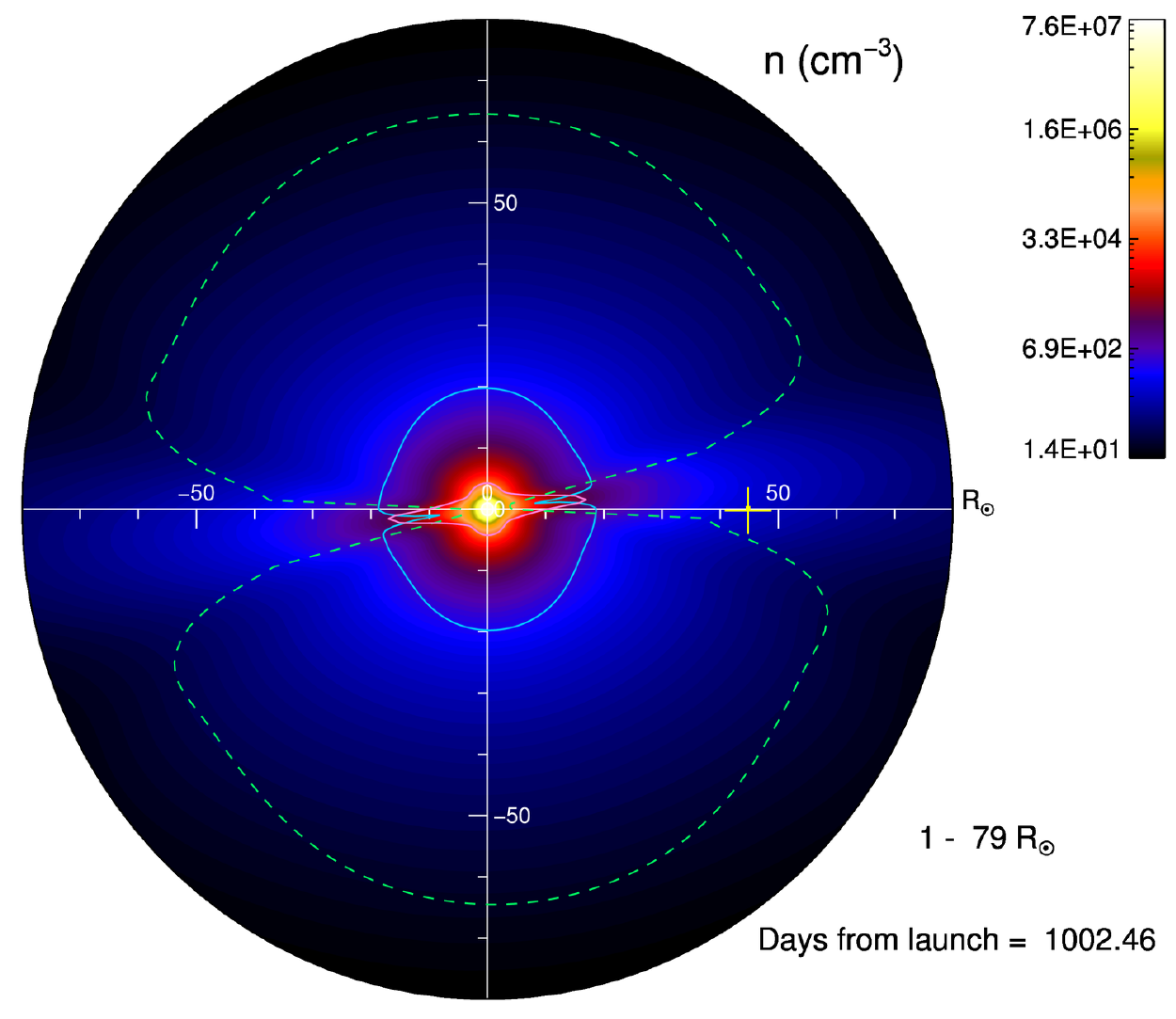}{0.45\textwidth}{(d)}
          }          
\caption{\psp crosssings of the critical surfaces are illustrated by a sequence of meridional planes that contain the spacecraft trajectory. The 8th orbit is depicted in a 10\degree~dipole simulation (Run I-C; see Figure \ref{fig:barplot2}(a)), representing solar-minimum conditions. The sonic, Alfv\'en, and first (proton+electron) beta unity surfaces are depicted as solid pink, solid blue, and dashed green curves, which are superposed on contours of proton density. The \psp position is at the center of the yellow `+' symbol. A video animation is available as Supplementary Material.}
\label{fig:pspmerid1}
\end{figure}

To produce an illustrative
comparison of the orbits and critical surfaces, 
we may choose to look at a sequence of (non-inertial) 
meridional planes that always contain the \psp 
orbit. 
In this frame the orientation of
the solar dipole field rotates at a non-constant angular frequency.  
Figure \ref{fig:pspmerid1} depicts such a sequence of meridional planes.
The MHD simulation used for this illustration employed a 10\degree~tilted dipole boundary condition (Run I-C), representing solar-minimum conditions likely to be sampled by the \psp in its early orbits. 
The position of \psp in each frame (during the 8th orbit; see Figure \ref{fig:barplot2}) is at the center of the yellow `+' symbol.
The
times are chosen to correspond to \psp passing over a critical surface. The plots are labeled by time 
measured in days-from-launch. \added{For these conditions, probably not unusual for early \psp orbits that occur during solar minimum, the spacecraft is often found skimming the edges of the \(\beta=1\) surface near the HCS. This may provide opportunities for \psp to study \(\beta\sim 1\) plasma for extended periods.} A video animation of these figures is available as 
Supplementary Material. An animation illustrating \psp crossings of critical surfaces in the final orbit, during solar-maximum conditions (Run II-A), is also available.

Another interesting 
way to visualize the relationship between 
the \psp orbit and the critical surfaces is 
to tally the time spent 
in each orbit within the \(\beta=1\) surface (henceforth \(\beta\) refers to the ``two-fluid'' plasma beta \(\beta_{p+e}\)), the Alfv\'en surface, and the sonic surface. \added{For the purposes of the present study, the initial  (``launch'') heliolongitude of the \psp is arbitrarily placed within the simulation. Rather than focus on a particular (arbitrary) trajectory, we consider \(\sim 100\) values of the initial longitude \(\phi_{\psp,0}\), ranging from  0\degree~to 359\degree, and perform an average over them. That is, for a given simulation run (that represents a particular type of solar conditions), we first compute the time spent within the critical surfaces during an orbit, for \textit{each} \psp trajectory defined by a value of \(\phi_{\psp,0}\). We then average these times over the different \(\phi_{\psp,0}\) to obtain a \textit{mean} number of hours within the surfaces, for each orbit. These results are presented in the following figures, discussed below.}

\begin{figure}
\gridline{\fig{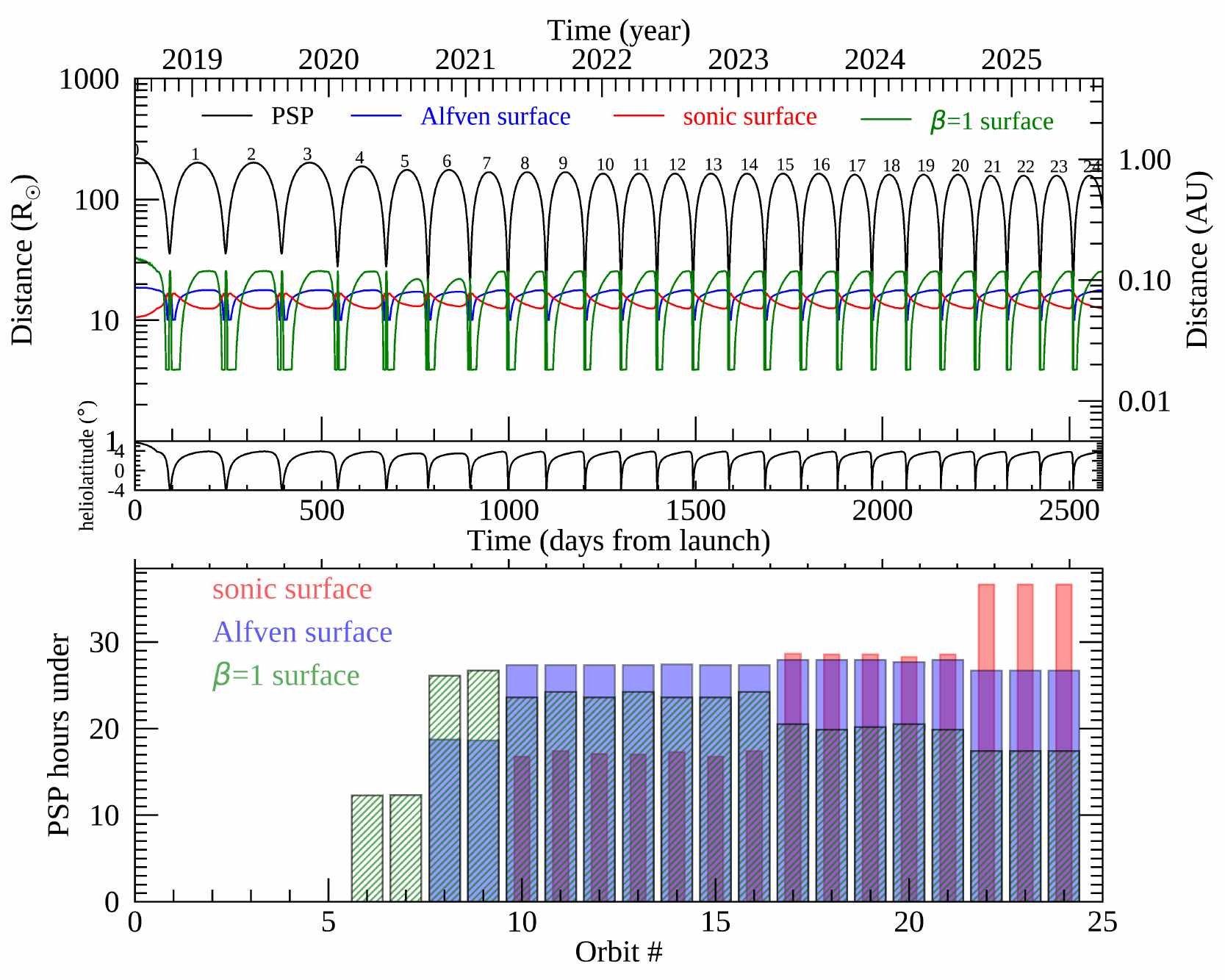}{0.65\textwidth}{(a)}
          }
\gridline{\fig{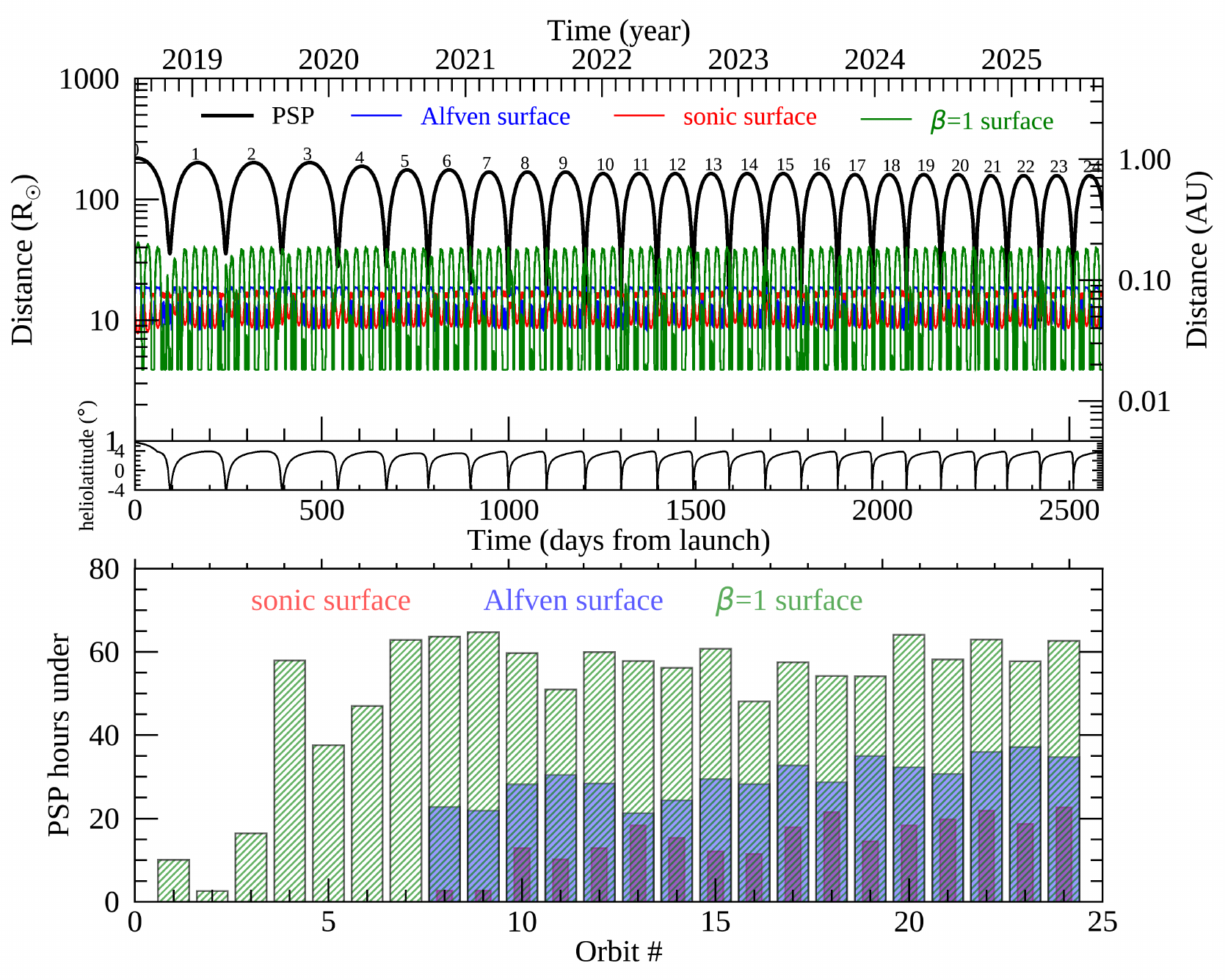}{0.65\textwidth}{(b)}
          }
\caption{\psp surface crossings from simulations with (a) 0\degree~ and (b) 5\degree~ dipole tilt. In each plot, the top section shows the radial and latitudinal position of the \psp for each orbit, and the radial position of the critical surfaces at the angular position of the \textit{PSP}. The bottom section shows the time spent by the \psp under each surface, per orbit. The striped green, lavender, and narrow red bars represent the \(\beta=1\), Alfv\'en, and sonic surfaces, respectively.}
\label{fig:barplot1}
\end{figure}

\begin{figure}
\gridline{\fig{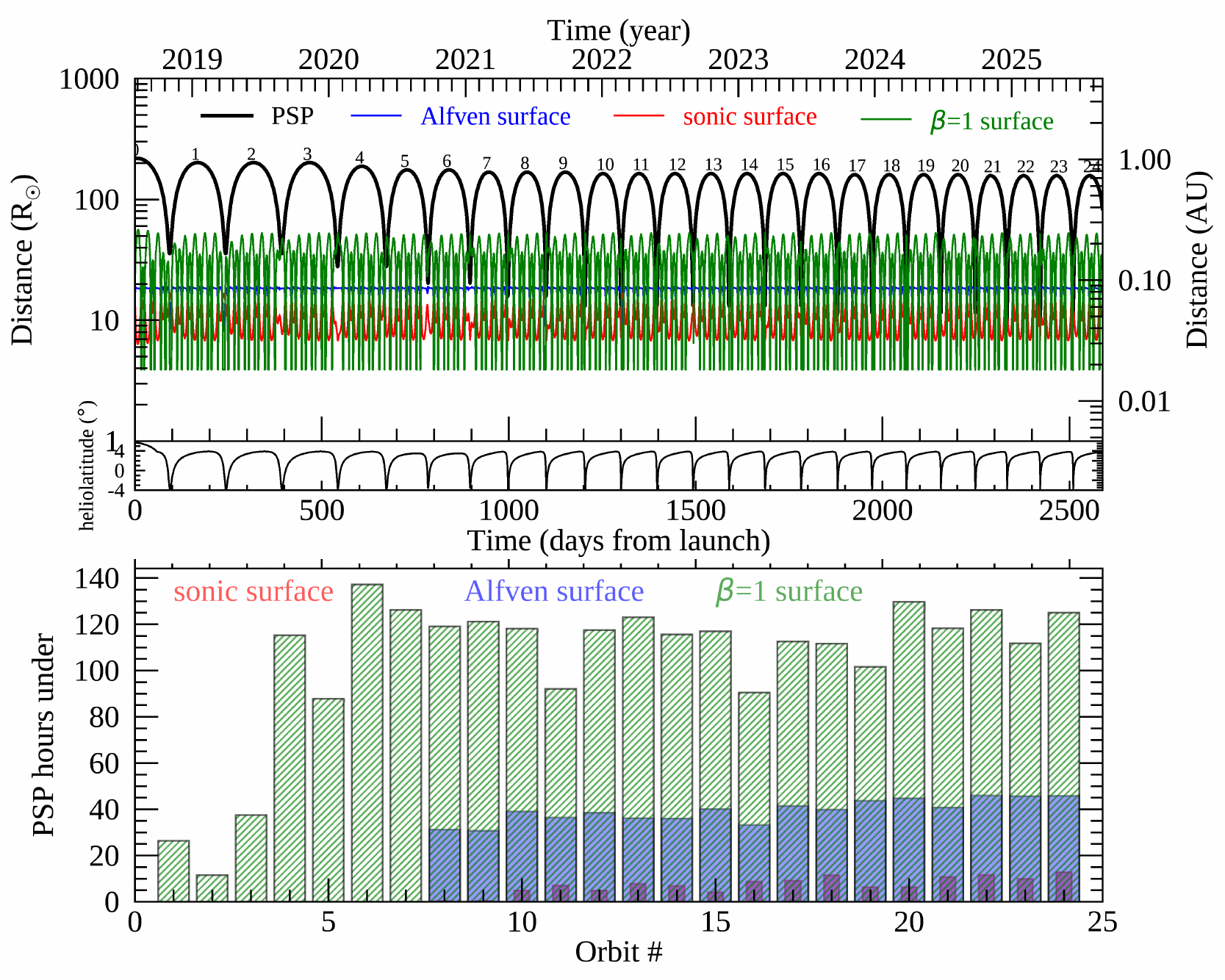}{0.65\textwidth}{(a)}
          }
\gridline{\fig{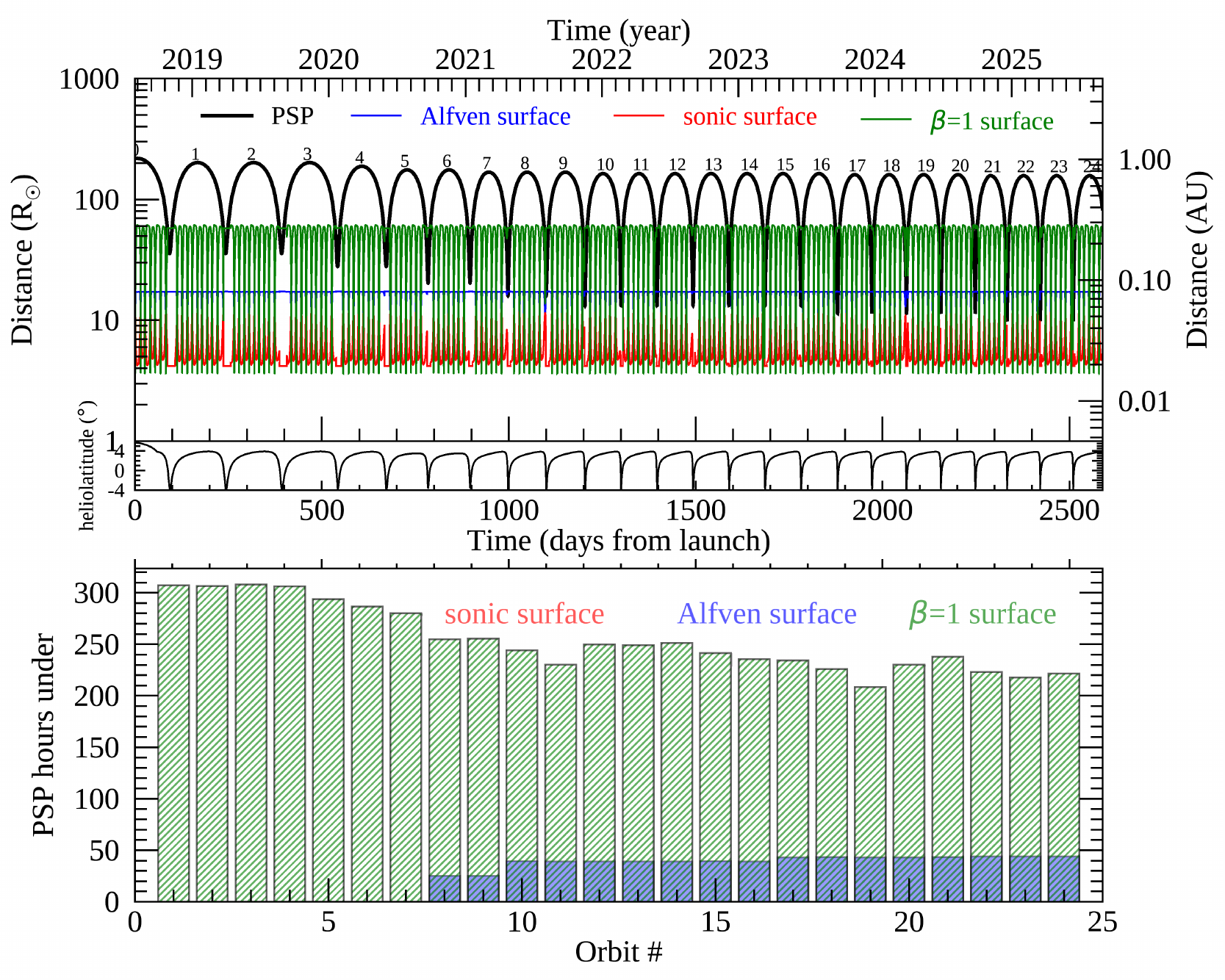}{0.65\textwidth}{(b)}
          }
\caption{\psp surface crossings from a simulation with (a) a 10\degree~and  (b) a 30\degree~dipole tilt. Further description follows Figure \ref{fig:barplot1}.}
\label{fig:barplot2}
\end{figure}

As a first example of this compilation,
Figure \ref{fig:barplot1}(a)
shows the residence time within each
of these regions, using the planned \psp orbits, 
for the case of a solar wind 
with untilted dipole boundary conditions. 
The upper section of the plot shows, as functions of 
time, the variation 
of orbital radial distances, as well as radial 
position of the critical surfaces at the angular position (heliolatitude and heliolongitude) of the \textit{PSP}, \added{for an arbitrary \(\phi_{\psp,0}\).} This directly illustrates \textit{PSP's} penetration of
the critical surfaces at various times.   

Referring to the lower section of Figure \ref{fig:barplot1}(a) that shows accumulated time \added{(averaged over \(\phi_{\psp,0}\))}
within critical surfaces, for each orbit, 
we see that,
beginning with orbit 8, this virtual \psp mission 
penetrates the Alfv\'en surface for 18 hours or more for all 
subsequent orbits to 25.  
Beginning with orbit 10, \psp spends between 
15 and 40 hours in each plotted orbit below the
predicted sonic surface. \replaced{There are 
no orbits falling  below
the \(\beta=1\) surface. This set of predictions is somewhat anomalous due 
to the lack of dipole tilt, so that the orbits almost always 
fall in the 
(artificially wide) high-\(\beta\) current sheet region.}{Due to the lack of dipole tilt an anomalous amount of time is spent in high-beta plasma, and the residence time below the \(\beta=1\) surface is supressed compared to subsequent cases. Recall also that the HCS in the simulation is artificially wide, and therefore the times spent within the \(\beta=1\) surface are likely to be underestimated, in particular for simulations with low dipole tilts.}\explain{We have changed an interpolation procedure used in the computation of the beta=1 surface, which produces a smoother and more gradual transition to the high-beta current sheet. Only Figure \ref{fig:barplot1}(a) is noticeably changed by this -- we now see a finite number of hours spent within the beta=1 surface for the untilted dipole case, in contrast to the previously submitted manuscript, in which the \psp did not spend any time under the beta=1 surface.} 

Figure \ref{fig:barplot1}(b)
shows a similar compilation 
done for a 5\degree~dipole-tilt run. 
We can see now, as would be expected,
that the encounters with critical surfaces 
have a strong dependence on the dipole tilt angle, which translates into the degree of latitudinal excursion of the HCS. 
In fact, for this case the critical surfaces are frequently 
seen at larger heliocentric distances, with significant 
consequences for the sub-critical-surface
residence times. The \(\beta=1\) surface is crossed relatively early, and from orbit 4 onwards \psp spends nearly 50 hours or more within it. Furthermore, for all orbits after 7, the \psp spends at least 20 hours within at least one of the 
critical surfaces.  These 20 to 40 hour periods
will represent opportunities for crucial observations. 
For instance, below the Alfv\'en surface the \psp might detect a large population of 
inward propagating Alfv\'en modes, and the enhanced turbulence seen in Figure 
\ref{fig:scintillation} could be detected in the trans-Alfv\'enic region.

Two more cases with dipole boundary conditions 
are shown in  
Figure \ref{fig:barplot2}, with tilt angles of 10\degree~and 30\degree. The results for a 60\degree~dipole run (not shown) are very similar to the 30\degree~case. It is apparent that the \(\beta=1\) surface is found at considerably larger radial distances as the tilt angle is increased. During solar maximum, the \psp is therefore likely to spend more than a hundred hours under the first beta unity surface per orbit. Furthermore, Figure \ref{fig:barplot2}(b) indicates that no time is spent within the sonic surface during any of the orbits in the 30\degree~dipole case. The reason for this can be understood from the discussion of Figure \ref{fig:lotova1997} -- Since the \psp trajectory stays within low heliolatitudes, it may be able to sample the extended portion of the sonic surface during solar minimum; However, during solar maximum the height of this surface is generally too low to be crossed at the latitudes sampled by the spacecraft (see also Figure \ref{fig:merid}(c)).
\subsection{What PSP will see: Magnetogram-based simulations}
\begin{figure}
\gridline{\fig{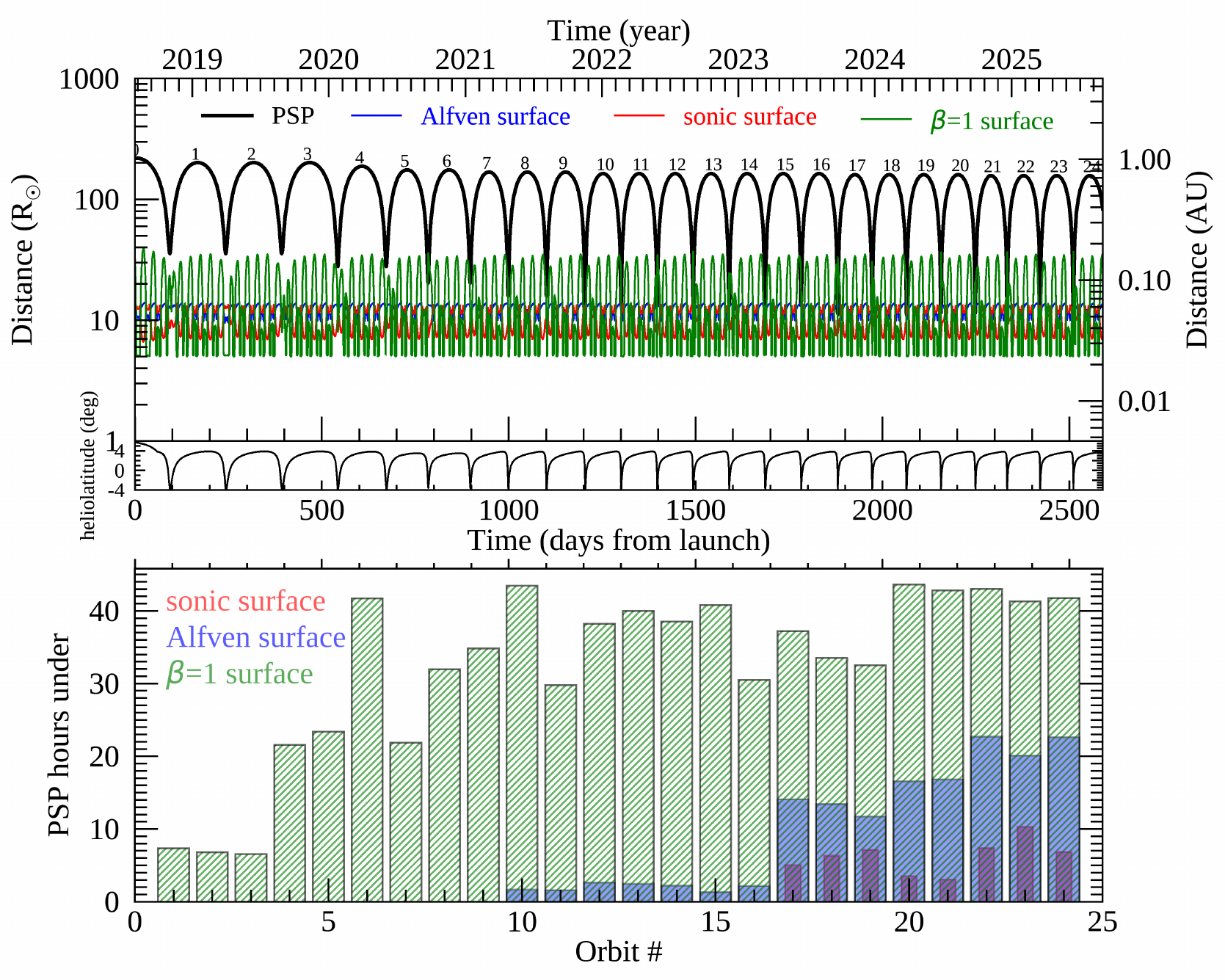}{0.65\textwidth}{(a)}
          }
\gridline{\fig{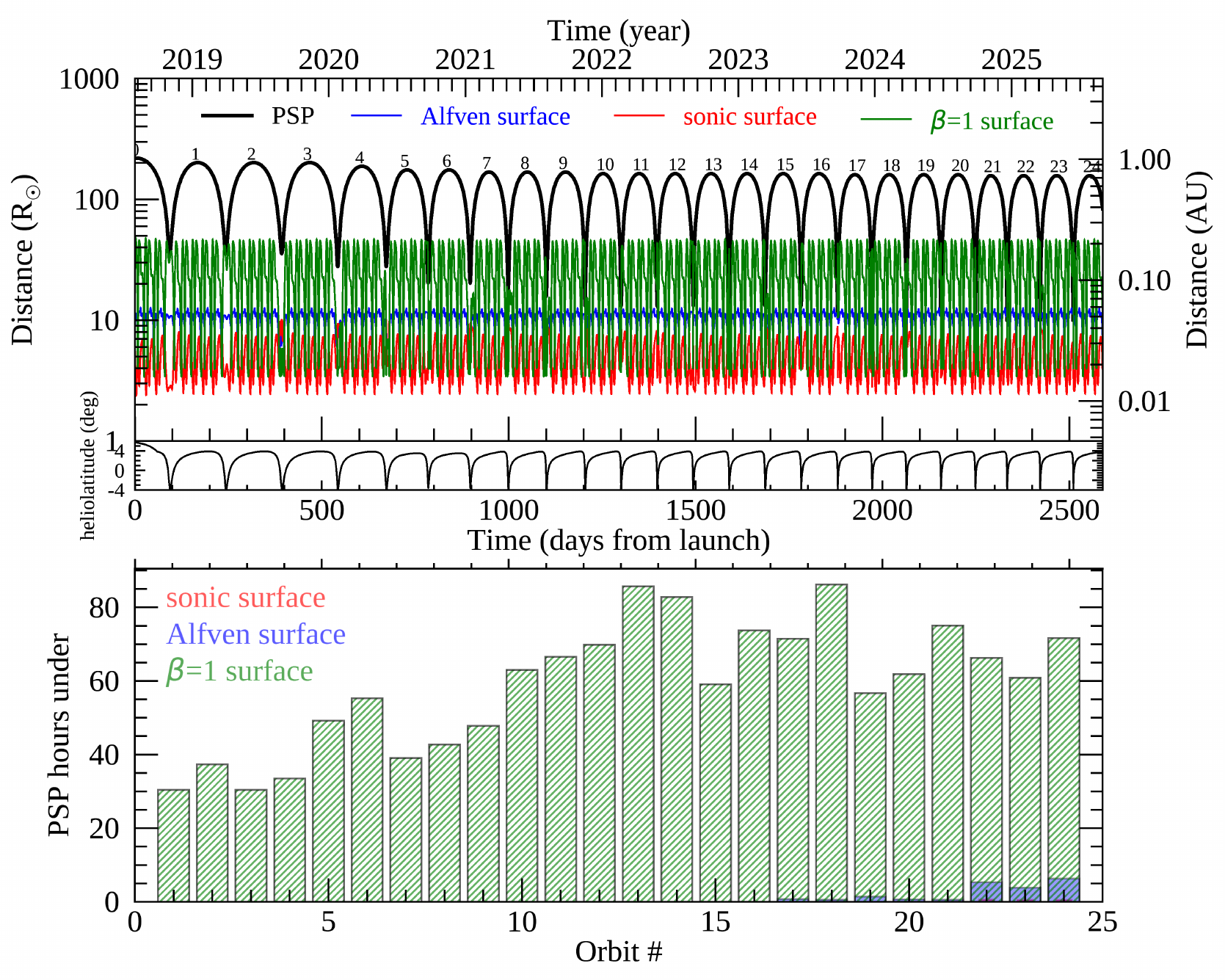}{0.65\textwidth}{(b)}
          }
\caption{\psp surface crossings for (a) a July 1994 (solar minimum) magnetogram run and (b) a July 1989 (solar maximum) magnetogram run. Further description follows Figure \ref{fig:barplot1}. A video animation of simulated \psp ``surface crossings'' in the solar maximum case is available as Supplementary Material.}
\label{fig:barplot3}
\end{figure}

Here we briefly show results for two cases in which 
the MHD simulation is driven by magnetograms:
one from solar minimum conditions (Carrington Rotation 1885, July 1994; Run II-B; Figure \ref{fig:barplot3}(a))
and another from solar maximum conditions (Carrington Rotation 1818, July 1989; Run II-A; Figure \ref{fig:barplot3}(b)). 
Examining the solar minimum case, one sees that the residence 
times within the Alfv\'en and sonic surfaces rarely, if ever, exceed twenty hours in a single orbit. 
Figure \ref{fig:barplot3}(b)
shows a solar maximum case employing a July 1989 magnetogram. 
The residence times under the \(\beta=1\) surface are
below 100 hours during any orbit. 
There are only a few orbits in which the Alfv\'en 
surface is encountered, and then for no more than 
about 10 hours in a single orbit. As indicated by Figure \ref{fig:barplot3}(b) (and Figure \ref{fig:barplot2}(b)), \psp crossings of the sonic surface are unlikely to occur during solar maximum. A video animation of simulated \psp ``surface crossings'' in the solar maximum case is available as Supplementary Material.

Compared with the dipole-based results (Figures \ref{fig:barplot1} and \ref{fig:barplot2}), the reduced time spent under the surfaces in Figure \ref{fig:barplot3} appears to be due to the rapid radial decay of the higher-order multipole magnetic fields that are implied by a complex magnetogram boundary condition \citep{reville2015ApJ798}. \added{It is also apparent that the \psp spends significantly fewer hours within the Alfv\'en surface in the solar maximum case (Figure \ref{fig:barplot3}(b)), compared to solar minimum (Figure \ref{fig:barplot3}(a)). The implied lowering of the Alfv\'en radius during solar maximum has been noted in other recent work as well \citep[e.g.,][]{pinto2011ApJ,pinto2017ApJ,perri2018JPP}.}

\added{While the decay of higher-order multipoles is a well-understood effect leading to radial reduction in fine-scale angular structure, this is somewhat offset by dynamical production of fine-scale structure in the corona and beyond. This effect is captured to a certain degree by existing models such as the present one but is also clearly limited by the ability to include fine-scale dynamics, that is, limited by spatial resolution of the numerics \citep{schmidt2015LRCA,miesch2015SSR194} as well as by the resolution of the boundary conditions (magnetogram resolution). Accordingly, more realistic global models of the solar atmosphere, like the real Sun, will include more will fine-scale structure at larger distances, and therefore the possibility of a larger number of brief passages through critical regions that we discuss here. In such cases interesting modifications might be expected to the depiction in Figure \ref{fig:barplot3} and to its comparison with Figure \ref{fig:barplot2}. The ``woodgrain'' structure obtained using high-resolution coronagraph imaging, as discussed by \cite{deforest2018ApJ}, hints at the appearance of such fine-resolution structuring in the real solar wind.}

%
%
\section{Conclusions and Discussion}\label{sec:disc}
We have shown here some detailed illustrative 
exercises in the use of a global heliospheric 
MHD code with turbulence modeling to simulate context that could be observed by 
the upcoming \textit{Parker Solar Probe} mission. 
We emphasize again 
that these results cannot be construed as predictions, 
since the boundary data employed are not only imprecise, but also are not 
appropriate to the conditions at the time 
when \psp will fly, except perhaps in a qualitative sense.  
Nevertheless it is interesting and even useful to explore the kind
of conditions that \psp might experience, an approach that we call 
\textit{context prediction}.

In this paper we have 
focused on ambient steady-state conditions in the solar wind, driven by boundary conditions 
that are simple untilted or tilted dipoles, or otherwise
magnetograms from previous solar minimum or solar maximum epochs.  
We note that a sensitive parameter is the total solar dipole strength, 
and we have used values commonly adopted in other work, 
which lead to agreement with 
near-Earth observations \citep{usmanov2014three,chhiber2018apjl,usmanov2018}, 
with the understanding that this value is actually not well constrained \citep{riley2014SoPh,usmanov2018}.

To summarize, the present results are of two major types: 
First, we 
find broad agreement in our study 
with the interpretation 
of existing remote sensing results, both from heliospheric 
imaging and from radio scintillation studies. 
Our results confirm the 
likely association of the region near the first outgoing $\beta=1$
surfaces  with morphological changes in the solar wind as observed 
in \textit{STEREO} imaging \citep{deforest2016ApJ828}.
Our global simulations also support the 
idea that a region near the critical Alfv\'en surfaces 
may be characterized by a local enhancement of turbulence 
levels, a feature that may have implications for additional 
heating and acceleration of the solar wind.
Second, the trajectory analyses show that the period of time that \psp is 
likely to spend inside the $\beta=1$, sonic, and Alfv\'en surfaces 
depends sensitively on the degree of solar activity and the tilt of 
the solar dipole and the location of the heliospheric current sheet.

Here we have provided a first set of such context predictions, 
emphasizing the possible range of positions of the sonic and 
Alfv\'enic critical surfaces, and the first plasma beta unity surface. 
The importance of these surfaces \citep[e.g.,][]{lotova1985AA150,deforest2016ApJ828,chhiber2018apjl}
lies in the fact that the physical character and 
conditions of the interplanetary medium are likely to be different on either side of these boundaries, which may in reality be very complex regions, or at
least corrugated surfaces. 
\textit{Parker Solar Probe} seeks to address questions such as the 
physical mechanisms that heat the corona and accelerate
the wind, and to reveal the structure of
the electromagnetic fields, plasma and energetic particles
in these very regions of the corona and wind.
Therefore, 
a baseline understanding the range of distances at which 
these regions might be encountered and crossed 
becomes quite important for anticipating 
what the mission is likely to measure, for how long, and
on which orbits.  
In a forthcoming 
paper 
we will continue these investigations, describing 
in some detail the turbulence properties that are expected
in the regions above and below the critical surfaces and along the \psp trajectory \citep[see also][]{cranmer2018RNAAS}, together with an evaluation of the validity of the Taylor hypothesis for \psp observations. 

\acknowledgments
We thank J. Kasper for useful discussions and the APL \psp project office
for providing the NASA SPICE kernel containing the \psp ephemeris. This research is supported in part by the NASA 
\textit{Parker Solar Probe} mission
through the IS\(\sun\)IS project and subcontract SUB0000165
from Princeton University to University of Delaware,
by the NASA HGC program grant NNX14AI63G, by the NASA
LWS program under grant NNX15AB88G, and by NASA HSR grants 80NSSC18K1210 and 80NSSC18K1648. The preparation of this article made use of the \href{http://adsabs.harvard.edu/}{SAO/NASA Astrophysics Data System (ADS)}.
%

%
%
\end{document}